\documentclass[USenglish,oneside,twocolumn]{article}

\usepackage[utf8]{inputenc}
\usepackage[big]{dgruyter_NEW}
\DOI{foobar}

\usepackage[final,inline,nomargin]{fixme}      
\usepackage{url}
\usepackage{amsmath}
\usepackage{amsfonts}
\usepackage[]{algorithm2e}
\usepackage{algpseudocode}

\makeatletter
\newcommand{\removelatexerror}{\let\@latex@error\@gobble}
\makeatother

\newlength\figureheight
\newlength\figurewidth
\newlength\smallfigureheight
\newlength\smallfigurewidth
\newlength\largefigureheight
\newlength\largefigurewidth
\usepackage{tikz}
\usepackage{filecontents}
\usepackage{pgfplots, pgfplotstable}
\usetikzlibrary{pgfplots.groupplots,shapes,snakes}
\pgfplotsset{compat=1.5,label style={text height=.5em,text depth=.1em},}
\usepgfplotslibrary{statistics}
\usetikzlibrary{external}

\usepackage{color}
\usepackage{soul}          
\usepackage{balance}
\usepackage{easy-todo}
\newcommand{\paragraphX}[1]{\vskip 4pt \noindent \textbf{#1} \hskip .05in}

\let\oldenumerate\enumerate
\renewcommand{\enumerate}{
  \oldenumerate
  \setlength{\topsep}{-7pt}
  \setlength{\leftmargin}{0pt}
  \setlength{\labelwidth}{0pt}
  \setlength{\itemsep}{2pt}
  \setlength{\parskip}{0pt}
  \setlength{\parsep}{0pt}
}

\usepackage{graphicx}
\usepackage{caption}
\usepackage{subcaption}
\usepackage[numbers]{natbib}
\graphicspath{{./Figures/}}

\begin{document}

  \author*[1]{Mohsen Imani}
  \author[2]{Mehrdad Amirabadi}
  \author[3]{Matthew Wright}
  \affil[1]{The University of Texas at Arlington, E-mail: mohsen.imani@mavs.uta.edu}
   \affil[2]{The University of Texas at Arlington, E-mail: mehrdad.amirabadi@mavs.uta.edu}
  \affil[3]{Rochester Institute of Technology, E-mail: mwright@rit.edu}

  \title{\huge Modified Relay Selection and Circuit Selection for Faster Tor}

  \runningtitle{Modified Relay Selection and Circuit Selection for Faster Tor}


\noindent \textbf{Abstract:} 
Users of the Tor anonymity system suffer from less-than-ideal performance, in part because circuit building and selection processes are not tuned for speed. In this paper, we examine both the process of selecting among pre-built circuits and the process of selecting the path of relays for use in building new circuits to improve performance while maintaining anonymity. First, we show that having three pre-built circuits available allows the Tor client to identify fast circuits and improves median time to first byte (TTFB) by 15\% over congestion-aware routing, the current state-of-the-art method. Second, we propose a new path selection algorithm that includes broad geographic location information together with bandwidth to reduce delays. In Shadow simulations, we find 20\% faster median TTFB and 11\% faster median total download times over congestion-aware routing for accessing webpage-sized objects. Our security evaluations show that this approach leads to better or equal security against a generic relay-level adversary compared to Tor, but increased vulnerability to targeted attacks. We explore this trade-off and find settings of our system that offer good performance, modestly better security against a generic adversary, and only slightly more vulnerability to a targeted adversary.


\journalname{}

  \journalyear{2018}
  \journalvolume{2018}
  \journalissue{x}
 
\maketitle
\section{Introduction}\label{Introduction}
Tor provides anonymity for millions of users around the world by routing their traffic over paths selected from approximately 7,000 volunteer-run relays.\footnote{\url{https://metrics.torproject.org/}, accessed October. 2017} Tor effectively hides the user among all the users, so having more users and more traffic enhances anonymity for all~\cite{acquisti03economics,usability:weis2006}. Unfortunately, Tor users often face large delays and long download times, which can discourage users and thereby reduce anonymity. In this paper, we examine two approaches to improve Tor performance and evaluate them in term of both performance and security.   

\paragraphX{Circuit Selection.}
The client's traffic in Tor goes through a three-hop encrypted channel, called a \textit{circuit}. When the user makes a request, such as for a webpage, Tor attaches the new 
stream (by opening a SOCKS connection) to a circuit. The Tor client builds circuits preemptively based on the client's use or immediately if there is no current circuit to handle the stream. 
Tor currently does not use any performance criteria in selecting a circuit. In this paper, 
we evaluate using the length of the circuits, their congestion, the Round Trip Time (RTT), or a combination of them in choosing a fast circuit. We also find that the number of available circuits in Tor is often small, between one and three circuits,
such that picking the best circuit for performance does not have much effect in practice. As the number of available circuits increases, the chance of finding a fast and high performance circuit should increase. To this end, for each circuit selection criteria we study, we evaluate the impact of more available circuits in terms of both performance and security.

\paragraphX{Relay Selection.}
For circuit selection to be effective, some of the available circuits must be reasonably high performing. To improve the chances of this, we modify the relay selection mechanism to build short and high-bandwidth circuits. Tor clients select paths in a way that balances traffic load among the relays according to their advertised bandwidths, but they do not make any consideration for the locations of relays relative to the clients, their destinations, or the other relays in the path. Paths can jump around the globe, which is {\em intuitively} good for anonymity but {\em measurably} bad for performance.

Prior work has examined improving path selection in Tor for better performance, considering factors such as bandwidth~\cite{tune_up}, congestion~\cite{congestion}, latency~\cite{Sherr09}, and location~\cite{lastor}. 

Wacek et al. performed a comprehensive study of path selection~\cite{empirical}, and they found that congestion-aware routing~\cite{congestion} offers the best combination of performance and anonymity among the tested approaches. They also found that approaches that emphasized latency but failed to consider bandwidth had poor performance, and they suggested that an approach that optimized both latency and bandwidth could do better than any of their tested approaches. In this paper, we take on this suggestion and explore designs that address both criteria. 

\vspace{0.3cm} \noindent We make the following contributions:
\begin{itemize}
\item We define nine circuit selection approaches using the geographical length, circuit delay, congestion, or a combination of these. We evaluate each of the approaches and compare them experimentally.

\item In our relay selection approach, {\em combined weighting}, we explore the design of a single weighting function that balances bandwidth and geographical inter-node distance. We examine the design issues in our approach and compare it with the state of the art. 

\item To prevent delays, it is important to build circuits in advance of their use~\cite{empirical}. Since we want to use destination location information to better inform our path-selection strategy, we build circuits in advance, using popular destinations as the end points. We then select from among these circuits base on our findings in circuit selection approaches.

\item We show the results of experiments on our approaches in Shadow, following the methodology of Jansen et al.~\cite{modeling}, and we examine a range of parameters.
We find a number of settings in our proposed methods that offer reasonable anonymity and significant performance improvements over congestion-aware routing, the current state-of-the-art in Tor path selection. In particular, our recommended approach provides a 20\% reduction in median time to first byte and a 11\% reduction in median time to last byte compared to congestion-aware routing.

\item We also measure the security of our approaches using Gini coefficient and entropy on first-and-last combinations, with the rate of path compromise in the presence of relay-level and AS-level adversaries, and in the presence of four targeted relay-level attacks. We find that both of our approaches provide anonymity in line with Tor at settings that also provide significant performance improvements. Our recommended approach has a slightly better Gini coefficient and entropy than Tor, with slightly fewer compromised paths against our attackers.
  
\end{itemize}

\section{Related work}\label{Related works}

Researchers have addressed Tor performance issues in a variety of ways, such as modifying circuit scheduling~\cite{scheduling}, congestion control~\cite{congestion}, traffic splitting~\cite{splitting}, and incentives to encourage users to offer their bandwidth~\cite{goldstar,braids}. In this section, we first briefly overview Tor's current path selection mechanism and then discuss the prior works on enhancing performance in Tor from circuit selection and relay selection point of view.

\paragraphX{Tor.}
Tor is a volunteer-based overlay network providing anonymity online. Details are available at \url{http://www.torproject.org/} and in the original design paper~\cite{tor}. In Tor, the choice of relays is governed a complex weighting function\footnote{Full details at \url{https://gitweb.torproject.org/torspec.git/tree/dir-spec.txt}.} that includes various considerations and the bandwidths of the relays. Weighting by bandwidths serves to balance load in the system, as relays have huge variance in advertised bandwidth, with the bottom quintile under 2 Mbps, a median of about 10 Mbps, and a maximum of 1 Gbps as of May 2016.\footnote{\url{http://metrics.torproject.org/}} 

\paragraphX{Circuit Selection.} 
Can et al.~\cite{scheduling} propose a circuit scheduling mechanism that gives high priority to interactive traffic over bulk traffic on the same connection. 
This circuit selection mechanism has been deployed
in Tor relays, but it has no impact on the client. Our circuit selection approaches are designed to improve the performance from the client side and are thus orthogonal to scheduling in the relays.

Wang et al. introduce \textit{node latency} as a parameter to measure a relay's congestion~\cite{congestion}. In their approach, {\em congestion-aware routing (CAR)}, the client calculates congestion delay using both active and opportunistic methods. It then uses the measured latency to avoid congested nodes during path selection and to avoid selecting congested paths. They use both {\em short-term} and {\em long-term} congestion in their work, where short-term congestion is caused by current traffic levels and long-term congestion is caused by the relay's bandwidth. Their results show improvement in quality of service and load balancing. In our evaluation of circuit selection methods, we also examine the use of congestion times and compare them with RTTs and circuit lengths. 

The current Tor client measures the Circuit Build Time (CBT), i.e. the time to construct the circuit, and uses this to discard slow circuits whose CBT is above a client-specific threshold.
Annessi and Schmiedecker~\cite{navigator} propose that Tor should use the circuit round trip times (RTTs) in eliminating slow circuits instead of CBTs. In this method, the circuit RTT is actively measured after the circuit is built, and if it is longer than a timeout the circuit is discarded from future uses. In their study, this provided only 3\% improvement in the time to download the first byte with mixed anonymity results. Our strategy in this paper is different from both approaches. Rather than examining circuits after their creation to discard them or keep them, we instead try to pick a high performing circuits in the first place.

\paragraphX{Relay Selection.} A number of improvements to Tor path selection have been investigated~\cite{tune_up,Sherr09,lastor}. Wacek et al. examine Tor path selection in a comprehensive study with experiments running many simultaneous clients~\cite{empirical}. They create a model of the Tor network to evaluate the recent published papers modifying path selection and show results for throughput, time to last byte (TTLB), and round-trip time (RTT).
They tested Tor, Snader/Borisov~\cite{tune_up}, Unweighted Tor, in which Tor relays get selected uniformly at random, Coordinate~\cite{Sherr09} in which path selection is based on estimated pair-wise latencies, LASTor~\cite{lastor}, and Congestion-Aware~\cite{congestion}. Their investigation shows that path selection algorithms that do not consider bandwidth as a factor in relay selection have poor performance. Congestion-aware had nearly the best performance in throughput and time-to-first-byte, plus it had anonymity approximately in line with Tor and significantly better than other high-performing algorithms. We thus select it for comparison in our work.

Improving performance can also affect attackers, potentially providing them better attacking opportunities and more accurate measurements. There are several attacks that use latency and throughput information to de-anonymize Tor users~\cite{congestion-longpaths, tissec-latency-leak, ccs2011-stealthy}. Geddes et al.~\cite{pets13-how-low} introduced a new class of attacks, called \textit{induced throttling}, that exploit performance-enhancing mechanisms to throttle and unthrottle a circuit and identify the user. They evaluated the vulnerability of performance improvements, such as congestion control and traffic admission control to these attacks, and they found that there are highly effective attacks that can uniquely identify users. While this does not directly affect our approaches, we recognize that there is generally a trade-off between anonymity and performance.


\section{Model and Goals}\label{model}

\subsection{Network Model}
Testing new path and circuit selection strategies on the live Tor network is challenging and could compromise users' anonymity or their harm their performance. 
We perform our simulations in Shadow~\cite{shadow,shadow_link}, a discrete-event simulator that runs the Tor code in a complete, but scaled-down, network. Shadow simulates the underlying network and it considers network attributes such as packet loss, bandwidth upstream and downstream, jitter, latency, and network edges. 
%
In our performance evaluations, we used a scaled-down model of Tor, which consists of 1100 clients and 220 relays; this scaled-down model was built based on the procedures of Jansen et al.~\cite{modeling} and measurements from the live Tor network (from July 2015).
In our security evaluations, we use a larger scaled-down model of 2127 relays with one client at a time.

\subsection{Attacker Model}
As with prior work in Tor performance~\cite{tune_up, Sherr09, lastor, empirical}, our attention is more on performance characteristics than on attacks. We only seek to validate that our approach does not significantly weaken the anonymity provided by Tor currently. We evaluate the security of our proposed mechanisms in terms of both relay-level and network-level adversaries. 

\paragraphX{Relay-Level Adversary Model.} In the relay-level adversary model, we assume that the adversary is running some Tor relays in the network with the goal of getting into the guard and exit positions of some circuits. An adversary in such a position can observe the entry and exit traffic and correlate them to link the clients to their destinations. To evaluate the security of our proposed circuit selection mechanism, we simulate our proposed method, CAR, and {\em vanilla} Tor in Shadow and randomly mark one of our guards and one of our exits as malicious relays. We then extract the streams and identify which ones were compromised. We repeat this process 10 times, and measure the compromise rates all over 10 repetitions. 

To evaluate the security of our proposed relay selection mechanism at the relay level, we first follow the approach of Wacek et al., who use the Gini coefficient and entropy as measures of the diversity of paths taken by each of their studied approaches~\cite{empirical}. 
%
We consider a high-bandwidth attacker who adds a modest number of high-bandwidth ORs into the Tor network. Since our path selection algorithm uses distance as well as bandwidth, leading to our path selection algorithms pick high-bandwidth ORs with short distance more often, this attacker is aimed at capturing a large number of circuits. We also consider four targeted attack strategies in which the attacker targets a specific client, a specific destination, a specific client and destination, or with no specific target. In all these strategies, the attacker places his relays in the target's exact location to have minimum distance and a high chance to be selected. Our targeted attack scenarios are thus worst cases.


\paragraphX{Network-Level Adversary Model.} The adversary can control some network components like ASes or IPXs. If the entry traffic and exit traffic of an anonymous connection traverse through the adversary's network components, the adversary observes both sides of the traffic and deanonymizes the clients. We evaluate the security of our circuits selection mechanisms and relays selection mechanisms in the network level.
In circuits selection and relay selection mechanisms, we simulate each of the proposed mechanisms in Shadow and extract the streams, including their paths. To determine the compromised streams, we use the algorithm proposed by Qiu and Gao~\cite{Qiu05aspath} to infer the AS paths on both the entry side of the circuit (between clients and guards) and and exit side (between exits and servers). Qiu and Gao's algorithm exploits known paths from BGP tables to improve the inferred paths. In measuring the compromise rates, we consider the possibility of an asymmetric traffic correlation attack that can happen between  {\em data} path and {\em ack} path, which is one of the RAPTOR attacks proposed by Sun et al.~\cite{raptor}.


\subsection{Design Goals}

We seek an algorithm that meets the following goals:
\begin{enumerate}
\item Interactive use like web browsing should be significantly faster than Tor and prior work. 
\item Performance for bulk downloads should not be significantly slowed compared to Tor.
\item Anonymity should be similar to what Tor currently provides against our selected attack models.
\item Usage should be fairly distributed among relays according to their available capacities. 
\item Clients should be able to select paths with little computational or other overhead.
\item Circuits should be available to the client for attaching streams to when needed.
\item We should avoid downloading large amounts of additional information from the directory servers. 
\end{enumerate}

We emphasize web traffic since delays in interactive use are more harmful to the user experience than delays in bulk downloads. We consider both response time, measured as time to first byte (TTFB), and total download time, measured as time to last byte (TTLB).

Note that we do not seek the optimal latency for circuits. Although having accurate latency information instead of geographic distance could further improve performance, the gains might be marginal given requirements for bandwidth and path diversity. Further, obtaining and distributing accurate pairwise latency information may be expensive due to the necessary measurements and directory server overhead.

\section{Circuit Selection}\label{circuit-selection}

When a Tor client issues a request, the new stream is handled by one of the available circuits. In this section we explain how Tor tries to provide some available circuits for new streams and how it attaches the streams to the circuits. Then we explain how the stream attachment can be improved by increasing the number of circuits and considering performance criteria in circuit selection.

\paragraphX{Pre-Built Circuits.}\label{Tor-preemptive}
As the user browses the Web with Tor,
the Tor client opens new circuits so that later streams can be attached to those circuits without delay. Since different exits support different sets of ports, the Tor client aims to keep open two circuits to cover any port that the user has used recently. In practice, one or two circuits are typically available at any given time.
 
\paragraphX{On-Demand Circuits.}\label{Tor-on-demand}
Sometimes the user's requested streams are not supported by current available circuits, or all available circuits are older than 10 minutes and considered {\em dirty}.
In this case, Tor builds a circuit for the unhandled stream and attaches the stream to this circuit. It is obvious that these streams experience more delay than streams using pre-built circuits due to the circuit built time.

\paragraphX{Tor Stream Attachment.}\label{Tor-stream-attachment}
When a new stream is created, the Tor client selects the most recently created circuit or creates a new circuit if needed and attaches the new stream to it. Then all communication on that stream, including DNS resolution, goes through the circuit.

\begin{figure*} [t]
	\centering
	\begin{subfigure}[ht]{0.56\textwidth}
		\includegraphics[width=85mm,height=21mm]{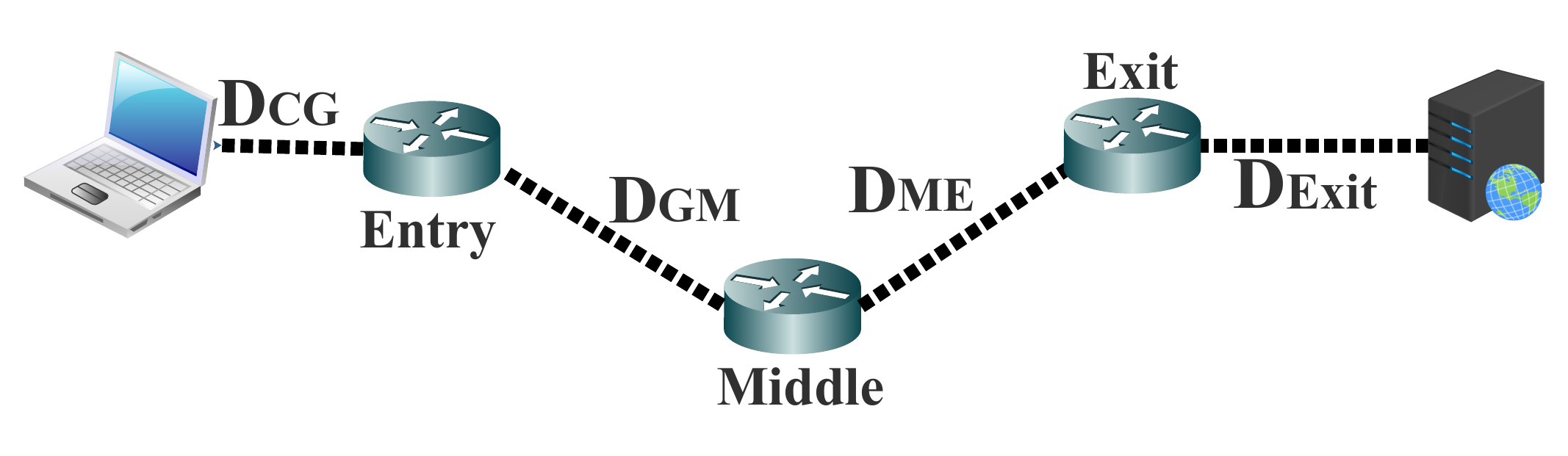}
		\caption{When the dest. IP is known}
		\label{fig:ip}
	\end{subfigure}
	\begin{subfigure}[ht]{0.34\textwidth}
		\includegraphics[width=55mm,height=21mm]{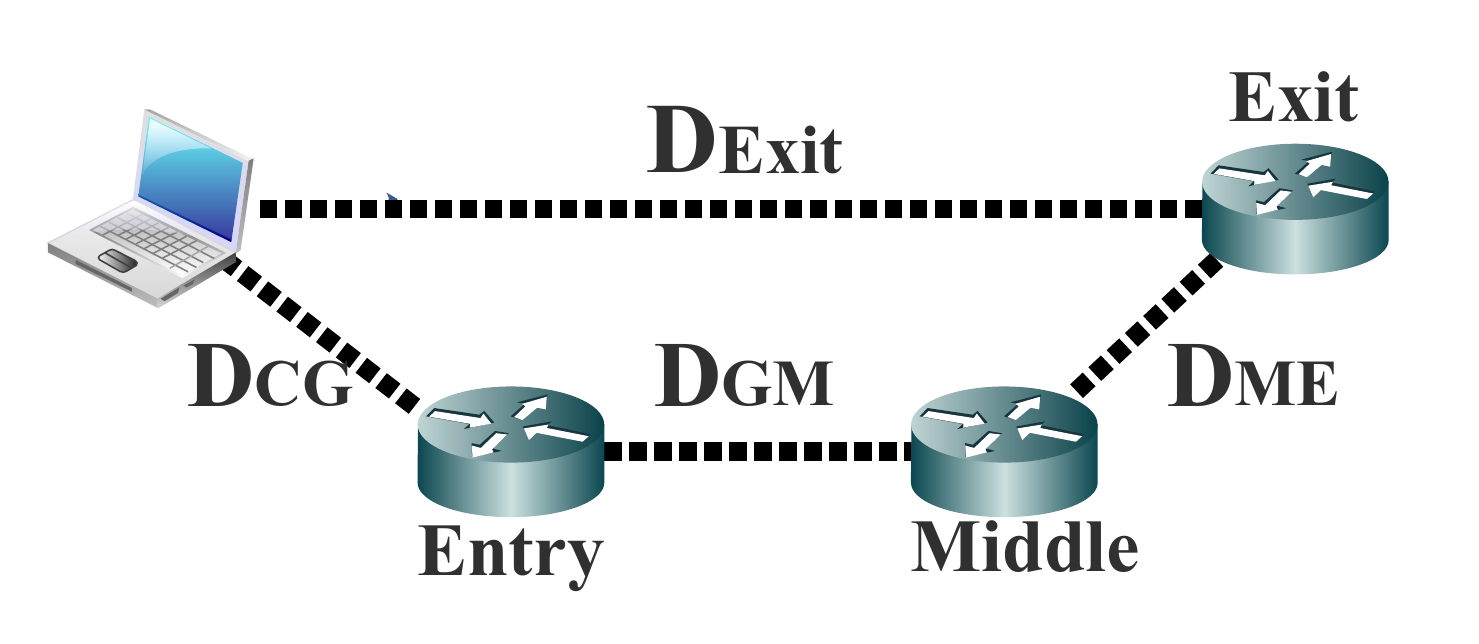}
		\caption{When the dest. IP is unknown}
		\label{fig:hostname}
	\end{subfigure}
	\caption{Distances used to compute circuit length.}
            \vskip -0.5cm
    \label{fig:stream_attachment}
\end{figure*}

\subsection{Performance in Circuit Selection}\label{Performance-circuit-selection}



The Tor client does not use performance as a criteria when selecting from available circuits for attaching a stream. Wang et al.~\cite{congestion} propose to use the least congested circuit, but there are several possible performance characteristics to use instead. Also, we know of no study testing the effect of changing number of circuits on performance-based selection.

To investigate the effect of circuit selection on Tor performance, we evaluate both the number of available circuits for the streams and the way we choose the best circuit among the available circuits. 
To set the number of circuits, we check once per second that there are at least $N$ circuits to support all recently used ports. If there are fewer than $N$ circuits, then we start building circuits to reach the threshold. We compare {\em vanilla} Tor, which typically offers one or two circuits, with making between three and five circuits available. Given some number of circuits, we can then use various methods to select the best one. We compare various combinations of geographic circuit length, congestion, and round trip time (RTT).

\paragraphX{Metrics.}
To find the total geographic circuit length (or simply {\em length}) $L$ between the client and destination, we compute:
\begin{equation}
L =  D_{CG} + D_{GM} + D_{ME} + D_{Exit}
\end{equation} 
\noindent $D_{CG}$, $D_{GM}$, $D_{ME}$ and $D_{Exit}$ are shown in Figure~\ref{fig:ip} for when the destination IP address is known and Figure~\ref{fig:hostname} for when the IP address is not yet known. 

We use the opportunistic circuit measurements and latency model proposed by Wang et al.~\cite{congestion} to measure the circuit round-trip times and congestion times. Congestion time $T_c$ is measured as:
\begin{equation}\label{equ:cong-time}
T_c =  RTT - RTT_{min}
\end{equation} 
where $RTT$ is the round-trip time and $RTT_{min}$ is the minimum RTT observed over that circuit. Wang et al.~\cite{congestion} showed that five measurements can effectively identify congested circuits. We thus measure and store the mean of the last five $T_c$ measurements as the congestion time of the circuit.


\subsection{Attaching Streams to Circuits}\label{attaching-streams}
We consider nine different methods in handling streams using circuit length, congestion, and RTT.

\begin{enumerate}
   \item \textit{Congestion Only}: Pick by lowest congestion time.

   \item \textit{Length Only}: Pick the shortest circuit.
   
   \item \textit{RTT Only}: Pick the circuit with the lowest RTT.
   
   \item \textit{Congestion then length}: Select the two lowest congestion times and pick the shorter one.
   
   \item \textit{RTT then length}: Select the two lowest RTTs and pick the shorter one.
   
   \item \textit{Length then congestion}: Select the two shortest circuits and pick the lower congestion time.
   
   \item \textit{Length then RTT}: Select the two shortest circuits and pick the lower RTT. 
 
   \item \textit{RTT then Congestion}: Select the two circuits with lowest RTT and pick the lower congestion time. 
   
   \item \textit{Congestion then RTT}: Select the two circuits with lowest congestion time and pick the lower RTT. 
   
\end{enumerate}


Since these circuit selection mechanisms are deterministic, given a set of candidate circuits, only one circuit from a set will be be used. These strategies will exploit the best circuit for the full 10-minute window that the circuit can be used. Since this means that the other circuits will go unused, we have the OP close any circuits that go unused for five minutes after their creation, leading to new circuits being opened. By itself, this might improve performance, as inferior circuits are closed in favor of untested circuits that may be better (or worse).


\section{Circuit Selection Performance}\label{experiment-circ-selection}
We now evaluate the nine methods of selecting circuits for stream attachment and compare them with Tor and CAR.

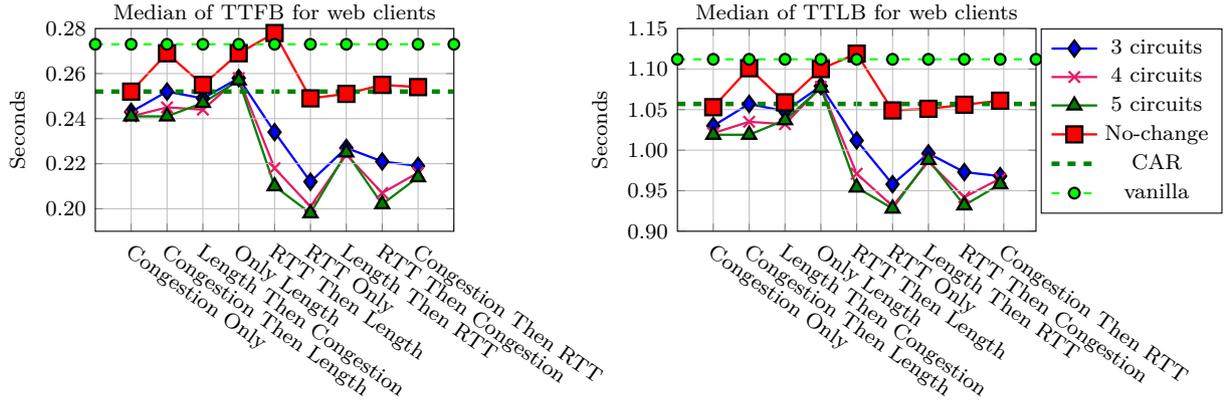
\begin{figure*}[t]
	\centering
        \pgfplotsset{footnotesize,samples=10}
\begin{center}%
\definecolor{color0}{rgb}{0.9,0.1,.4}
\begin{tikzpicture}
\begin{axis}[
title={Median of TTFB for web clients},
ylabel={Seconds},
xmin=0, xmax=10,
ymin=0.19, ymax=0.28,
axis on top,
width=0.9\figurewidth,
height=0.85\figureheight,
xtick={1,2,3,4,5,6,7,8,9,10},
xmajorgrids,
ymajorgrids,
xticklabels={Congestion Only,Congestion Then Length,Length Then Congestion,Only Length,RTT Then Length,RTT Only,Length Then RTT,RTT Then Congestion,Congestion Then RTT},
ytick={0.18,0.2,0.22,0.24,0.26,0.28,0.3},
yticklabels={,$0.20$,$0.22$,$0.24$,$0.26$,$0.28$,},
x tick label style={rotate=-35, anchor=north west, inner sep=1mm},
]
\addplot [line width = 0.3mm, blue, mark size=3, mark options={solid,draw=black}, mark=diamond*]
table {%
1 0.243
2 0.252
3 0.249
4 0.258
5 0.234
6 0.212
7 0.227
8 0.221
9 0.219
};
\addplot [line width = 0.3mm, color0, mark=x, mark size=3, mark options={solid}]
table {%
1 0.241
2 0.245
3 0.244
4 0.258
5 0.218
6 0.201
7 0.224
8 0.207
9 0.216
};
\addplot [line width = 0.3mm, green!50.0!black, mark=triangle*, mark size=3, mark options={solid,draw=black}]
table {%
1 0.241
2 0.241
3 0.247
4 0.257
5 0.21
6 0.198
7 0.225
8 0.202
9 0.214
};
\addplot [line width = 0.3mm, red, mark=square*, mark size=3, mark options={solid,draw=black}]
table {%
1 0.252
2 0.269
3 0.255
4 0.269
5 0.278
6 0.249
7 0.251
8 0.255
9 0.254
};
\addplot [line width = 0.6mm, green!50.0!black, dashed]
table {%
0 0.252
10 0.252
};
\addplot [line width = 0.3mm,thick, green, dashed, mark=*, mark size=2,mark options={solid,draw=black}]
table {%
0 0.273
1 0.273
2 0.273
3 0.273
4 0.273
5 0.273
6 0.273
7 0.273
8 0.273
9 0.273
10 0.273
};
\end{axis}
\end{tikzpicture}
\qquad 
\hspace{-2.0cm}
\qquad 
\begin{tikzpicture}
\begin{axis}[
title={Median of TTLB for web clients},
ylabel={Seconds},
xmin=0, xmax=10,
ymin=0.9, ymax=1.15,
axis on top,
xmajorgrids,
ymajorgrids,
width=0.9\figurewidth,
height=0.85\figureheight,
xtick={1,2,3,4,5,6,7,8,9.10},
xticklabels={Congestion Only,Congestion Then Length,Length Then Congestion,Only Length,RTT Then Length,RTT Only,Length Then RTT, RTT Then Congestion, Congestion Then RTT},
ytick={0.9,0.95,1,1.05,1.1,1.15},
x tick label style={rotate=-35, anchor=north west, inner sep=1mm},
yticklabels={$0.90$,$0.95$,$1.00$,$1.05$,$1.10$,$1.15$},
legend style={at={(1.52,1.0)}, anchor=north east},
legend entries={{3 circuits},{4 circuits},{5 circuits},{No-change},{CAR},{vanilla}},
]
\addplot [line width = 0.3mm, blue, mark size=3, mark options={solid,draw=black}, mark=diamond*]
table {%
1 1.03
2 1.057
3 1.048
4 1.079
5 1.012
6 0.958
7 0.996
8 0.973
9 0.968
};
\addplot [line width = 0.3mm, color0, mark=x, mark size=3, mark options={solid}]
table {%
1 1.021
2 1.035
3 1.032
4 1.078
5 0.971
6 0.931
7 0.986
8 0.942
9 0.965
};
\addplot [line width = 0.3mm, green!50.0!black, mark=triangle*, mark size=3, mark options={solid,draw=black}]
table {%
1 1.019
2 1.019
3 1.037
4 1.077
5 0.954
6 0.928
7 0.988
8 0.932
9 0.958
};
\addplot [line width = 0.3mm, red, mark=square*, mark size=3, mark options={solid,draw=black}]
table {%
1 1.053
2 1.101
3 1.059
4 1.1
5 1.119
6 1.049
7 1.051
8 1.056
9 1.061
};
\addplot [line width = 0.6mm, green!50.0!black, dashed]
table {%
0 1.057
10 1.057
};
\addplot [line width = 0.3mm,thick, green, dashed, mark=*, mark size=2,mark options={solid,draw=black}]
table {%
0 1.112
1 1.112
2 1.112
3 1.112
4 1.112
5 1.112
6 1.112
7 1.112
8 1.112
9 1.112
10 1.112
};
\end{axis}
\end{tikzpicture}
\end{center}
        \vskip -0.6cm
        \caption{{\bf Circuit Selection:} TTLB and TTFB for web clients.\\}
        \label{fig:circ-perf}
\end{figure*}


\subsection{Network Configuration}\label{Configuration}
We largely follow the experimental procedures suggested by Jansen et al.~\cite{modeling} and describe them here in brief. Shadow runs actual Tor code for accurate modeling; we used Tor version 0.2.5.12, modifying it as necessary to implement our methods and CAR. To generate a realistic Tor network topology, Shadow comes with topology generation tools that model a private Tor network based on a validated research study~\cite{modeling}. We used these tools and data from the Tor metrics portal to generate our private Tor network. Our Tor network includes 1100 clients, 220 Tor relays (52 exit relays, including exit-guard relays, and 49 guard relays), three directory authorities, and 220 HTTP destination servers. 

Shadow uses an underlying topology that models the Internet. The default topology shipped with Shadow is very small, consisting of only 183 vertices and 17,000 edges, and is not a good representation of the Internet.
For all the simulations in this paper we used the Internet topology used by Jansen at al.~\cite{kist}. This topology is provided by techniques from recent research in modeling Tor typologies~\cite{modeling,johnson:ccs13}, traceroute data from CAIDA~\cite{caida}, data from the Tor Metrics Portal~\cite{metrics-portal} and Alexa~\cite{alexa}, and it includes 699,029 vertices and 1,338,590 edges.
In our simulation, we tried different ratios of clients to relays, i.e. different congestion levels, and different average packet loss rates in the Internet topology and compared our results with Torperf data~\cite{metrics-portal}. We found that a clients-to-relays ratio of 5:1 with 0.0025\% packet loss provides us comparable results on TTFB and TTLB
with Torperf data.

Our clients run Tor code in client-only mode and are distributed around the world in line with Tor usage statistics. We have two types of clients in our experiments: web clients and bulk clients. The 900 web clients download 320 KiB of data and simulate web surfing behavior by waiting between 1 to 20 seconds uniformly at random before starting the next download. The 100 bulk clients download 5 MiB of data without pausing between the end of a download and starting the next one. This methodology is proposed in~\cite{modeling}. 

\subsection{CAR: Congestion-Aware Routing}\label{comparison}
To compare our methods, we also simulated CAR, the circuit selection technique of Wang et al.~\cite{congestion}. They proposed opportunistic and active probing techniques to measure congestion times, which are obtained by RTTs in Equation~\ref{equ:cong-time}, and use these measurements to mitigate congestion using both an instant response for temporary congestion and a long-term response for low-bandwidth conditions. In our simulation, we follow the method of Wacek et al.~\cite{empirical}, who also simulated CAR and ignored the long-term response due to its small impact on performance. 

When attaching streams to circuits in CAR, we randomly select three circuits from the circuit list and pick the one that has the smallest mean congestion time from the five most recent measurements. If the mean of last five congestion times is more than 0.5 seconds for a circuit, we stop using the circuit for new streams.  

\subsection{Performance Results}\label{Performance results}
Figure~\ref{fig:circ-perf} shows the median time-to-first-byte (TTFB) and time-to-last byte (TTLB) for web clients. {\em vanilla} represents unmodified Tor circuit selection, and {\em No change} represents the case we do not modify the number of circuits from Tor, which typically has one or two circuits available at a time.

As shown in Figure ~\ref{fig:circ-perf}, RTT is best criterion to choose the circuit, with {\em RTT Only} as the best method overall. {\em RTT Only} has 15\% lower TTFB than CAR (22\% lower than {\em vanilla}) for three circuits and 22\% lower TTFB than CAR (27\% lower than {\em vanilla}) for five circuits.
{\em RTT Only} also has 9\% lower TTLB than CAR (13.8\% lower than {\em vanilla}) for three circuits and 12\% lower TTLB than CAR (16\% lower than {\em vanilla} for five circuits.
We speculate that RTT is the best criteria because it effectively captures both propagation delays and congestion time (queuing delays and transmission delays). 

As expected, CAR is better than {\em vanilla}, and {\em Congestion Only} with {\em no change} in the number of circuits performs the same as CAR, as both use the same criteria. Length turns out to be less effective compared to RTT or congestion times, particularly when the number of circuits is small. We note that {\em Length Then RTT} performs fairly well for three or more circuits. Length may be suitable for gauging broad performance information, such as comparing a circuit with multiple intercontinental hops to one with no such hops, but poor at predicting the best circuit otherwise. 

The TTFB and TTLB for {\em RTT Then Congestion} are slightly better than {\em Congestion Then RTT}, which indicates that RTT can narrow down candidate circuits better than congestion times.  Results of both {\em RTT Then Congestion} and {Congestion Then RTT} are worse than {\em RTT Only} and show that mixing congestion time with RTTs will not provide better performance than using RTT by itself.

\subsection{Circuit Creation Analysis}\label{load}

\begin{figure*}[h]
	\centering
        \input{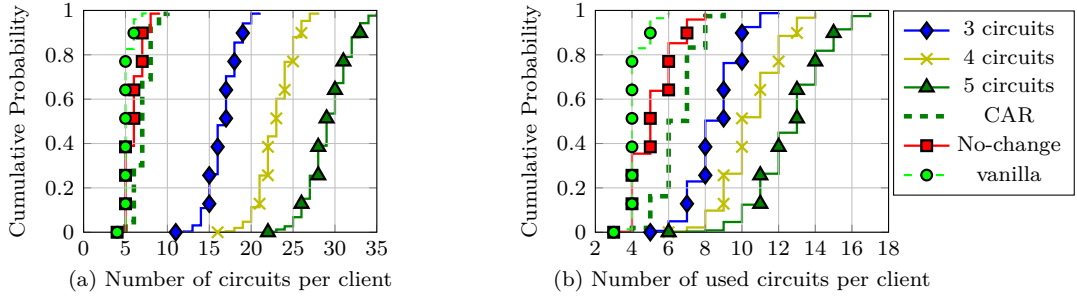}
        \caption{CDF of the number of circuits created (a) and used (b) for web clients.}
        \label{fig:cdf-circs}
\end{figure*}

In our circuit selection strategies we build more circuits than Tor normal behavior, so it is important to understand the load this imposes on the network. Unfortunately, Shadow does not provide results regarding the load on nodes. To estimate changes in load, we compare the strategies based on the number of created circuits and used circuits. To see how many circuits our clients build, we simulate the circuit selection strategies in Shadow for one hour of simulated time, which leads to about 40 minutes of activity after 20 minutes of initialization. We extract the number of general-purpose circuits built by our web clients. Figure~\ref{fig:cdf-circs} shows the CDF of created {\em general purpose} circuits and the CDF of used circuits, the circuits actually being used. We see that web clients in {\em vanilla}, CAR, {\em RTT Only} with the same number of circuits as Tor, and {\em RTT Only} with 3-5 circuits. 
The median of created circuits in {\em RTT Only} is 17, 23, and 29 circuits as we increase the number of circuits from 3 to 5.
{\em RTT Only} leads to building so many circuits due to proactively checking every second that there are $N$ circuits available for each recently used port, plus killing unused circuits after five minutes.
Fig.~\ref{fig:cdf-circs}.b shows how many of these created circuits have been used in transferring data. The median for used circuits in {\em vanilla}, CAR, and {\em RTT Only} with no change in the number of circuits is around four circuits, which means that they use all the circuits created and attach some stream to them. The median in {\em RTT Only} is 8, 10, and 13 circuits, respectively. 

\section{Security Analysis}\label{security-circ-selection}

In this section we examine the security of these circuit selection strategies, considering both relay-level and network-level adversaries. Our performance results showed that {\em RTT Only} outperforms all the other circuit selection strategies and CAR. Therefore, in this section, we focus on the security analysis of {\em RTT Only}.

\subsection{Relay-Level Adversary}\label{relay-level-circ-selection}
In relay-level adversary model, we assume that the adversary runs both guard and exit relays in the hope that his relays simultaneously occupy the guard and exit positions in some circuits. If the adversary can sit on the exit and guard position on a circuit, he can apply a traffic correlation attack and link the client to her destinations. These circuits and streams attached to them are called compromised circuits and compromised streams, respectively.   
To analyze the security of the {\em RTT Only} strategy, we need to have access to RTTs (which include propagation delays, queuing delays, and transmission delays), which means we need to simulate a whole network. To do this, we use again Shadow to simulate the Tor network, and we use the same Tor network configuration as our performance evaluations in Section \ref{Configuration}, which consists of 52 exit relays, including exit-guard relays, and 49 guard relays. 

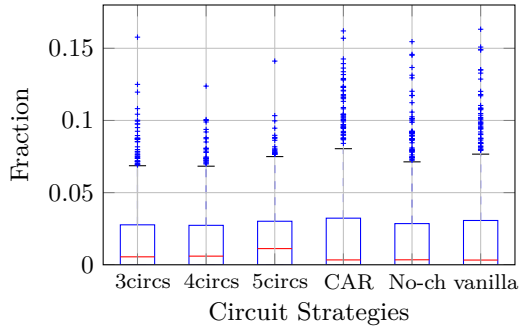
\begin{figure}[t!]
	\centering
%
%
%
\begin{tikzpicture}

\begin{axis}[
xlabel={Circuit Strategies},
ylabel={Fraction},
xmin=0.5, xmax=6.5,
ymin=0, ymax=0.18,
axis on top,
yticklabel style={/pgf/number format/fixed},  
width=\figurewidth,
height=\figureheight,
xtick={1,2,3,4,5,6},
xticklabels={3circs,4circs,5circs,CAR,No-ch,vanilla},
xticklabel style = {font=\small,xshift=0.5ex},
xmajorgrids,
ymajorgrids
]
\addplot [blue, dashed]
table {%
1 0
1 0
};
\addplot [blue, dashed]
table {%
1 0.0276955522440836
1 0.0686427457098284
};
\addplot [black]
table {%
0.875 0.0686427457098284
1.125 0.0686427457098284
};
\addplot [black]
table {%
0.875 0
1.125 0
};
\addplot [blue]
table {%
0.75 0
1.25 0
1.25 0.0276955522440836
0.75 0.0276955522440836
0.75 0
};
\addplot [black]
table {%
1 0.0276955522440836
1 0.0692431561996779
};
\addplot [red]
table {%
0.75 0.0055211147393569
1.25 0.0055211147393569
};
\addplot [blue, mark=+, mark size=1, mark options={solid}, only marks]
table {%
1 0.0692431561996779
1 0.0693481276005548
1 0.0697674418604651
1 0.07
1 0.0705679862306368
1 0.0716417910447761
1 0.0716510903426791
1 0.0717948717948718
1 0.0720720720720721
1 0.0736842105263158
1 0.0738161559888579
1 0.0744186046511628
1 0.0749279538904899
1 0.0752840909090909
1 0.0762362637362637
1 0.0762564991334489
1 0.0773638968481375
1 0.0782208588957055
1 0.0785714285714286
1 0.0798258345428157
1 0.0817518248175182
1 0.0852130325814536
1 0.0854037267080745
1 0.0868306801736614
1 0.0868878357030016
1 0.0873015873015873
1 0.0875486381322957
1 0.0876897133220911
1 0.0877437325905292
1 0.0895522388059701
1 0.092511013215859
1 0.0950468540829987
1 0.097542242703533
1 0.0976253298153034
1 0.0995405819295559
1 0.100149476831091
1 0.104321907600596
1 0.108191653786708
1 0.119631901840491
1 0.125
1 0.157718120805369
};
\addplot [blue, dashed]
table {%
2 0
2 0
};
\addplot [blue, dashed]
table {%
2 0.0273411183471817
2 0.0683139534883721
};
\addplot [black]
table {%
1.875 0.0683139534883721
2.125 0.0683139534883721
};
\addplot [black]
table {%
1.875 0
2.125 0
};
\addplot [blue]
table {%
1.75 0
2.25 0
2.25 0.0273411183471817
1.75 0.0273411183471817
1.75 0
};
\addplot [red]
table {%
1.75 0.00599721718333097
2.25 0.00599721718333097
};
\addplot [blue, mark=+, mark size=1, mark options={solid}, only marks]
table {%
2 0.0698412698412698
2 0.0705218617771509
2 0.0706666666666667
2 0.0717905405405405
2 0.0719874804381847
2 0.0725806451612903
2 0.072700296735905
2 0.0736434108527132
2 0.0738161559888579
2 0.0739299610894942
2 0.0739644970414201
2 0.0747217806041336
2 0.0778546712802768
2 0.0782608695652174
2 0.079646017699115
2 0.0805921052631579
2 0.0808080808080808
2 0.0850767085076709
2 0.0873460246360582
2 0.0881118881118881
2 0.0920245398773006
2 0.0937931034482759
2 0.0989505247376312
2 0.1
2 0.10062893081761
2 0.12379421221865
};
\addplot [blue, dashed]
table {%
3 0
3 0
};
\addplot [blue, dashed]
table {%
3 0.0301726011912526
3 0.0749665327978581
};
\addplot [black]
table {%
2.875 0.0749665327978581
3.125 0.0749665327978581
};
\addplot [black]
table {%
2.875 0
3.125 0
};
\addplot [blue]
table {%
2.75 0
3.25 0
3.25 0.0301726011912526
2.75 0.0301726011912526
2.75 0
};
\addplot [red]
table {%
2.75 0.0112179775280899
3.25 0.0112179775280899
};
\addplot [blue, mark=+, mark size=1, mark options={solid}, only marks]
table {%
3 0.0765625
3 0.0767004341534009
3 0.076797385620915
3 0.0769230769230769
3 0.0774299835255354
3 0.0775969962453066
3 0.0780234070221066
3 0.0783645655877342
3 0.0791268758526603
3 0.0801232665639445
3 0.0802292263610315
3 0.0817518248175182
3 0.0862800565770863
3 0.0876288659793814
3 0.0882352941176471
3 0.0897009966777409
3 0.0946502057613169
3 0.0997150997150997
3 0.103343465045593
3 0.141129032258065
};
\addplot [blue, dashed]
table {%
4 0
4 0
};
\addplot [blue, dashed]
table {%
4 0.0323059979667909
4 0.0804438280166436
};
\addplot [black]
table {%
3.875 0.0804438280166436
4.125 0.0804438280166436
};
\addplot [black]
table {%
3.875 0
4.125 0
};
\addplot [blue]
table {%
3.75 0
4.25 0
4.25 0.0323059979667909
3.75 0.0323059979667909
3.75 0
};
\addplot [red]
table {%
3.75 0.00338251272373763
4.25 0.00338251272373763
};
\addplot [blue, mark=+, mark size=1, mark options={solid}, only marks]
table {%
4 0.0841121495327103
4 0.0866141732283465
4 0.0871794871794872
4 0.0873146622734761
4 0.0875816993464052
4 0.0879478827361563
4 0.0898021308980213
4 0.0899854862119013
4 0.0905292479108635
4 0.0916530278232406
4 0.0924499229583975
4 0.0944
4 0.0949367088607595
4 0.0953101361573374
4 0.0953177257525084
4 0.0954773869346734
4 0.0961262553802009
4 0.0969101123595506
4 0.0970588235294118
4 0.0970724191063174
4 0.099023709902371
4 0.1
4 0.100671140939597
4 0.101108033240997
4 0.102240896358543
4 0.102564102564103
4 0.103618421052632
4 0.10413694721826
4 0.104144527098831
4 0.104477611940299
4 0.10806697108067
4 0.108965517241379
4 0.113237639553429
4 0.115079365079365
4 0.115789473684211
4 0.116219667943806
4 0.117154811715481
4 0.11791730474732
4 0.120443740095087
4 0.120915032679739
4 0.122861586314152
4 0.122962962962963
4 0.12338593974175
4 0.123664122137405
4 0.128099173553719
4 0.12824427480916
4 0.129084967320261
4 0.131795716639209
4 0.133757961783439
4 0.136298421807747
4 0.139380530973451
4 0.142614601018676
4 0.156976744186047
4 0.1620294599018
};
\addplot [blue, dashed]
table {%
5 0
5 0
};
\addplot [blue, dashed]
table {%
5 0.0285444744990429
5 0.0713245997088792
};
\addplot [black]
table {%
4.875 0.0713245997088792
5.125 0.0713245997088792
};
\addplot [black]
table {%
4.875 0
5.125 0
};
\addplot [blue]
table {%
4.75 0
5.25 0
5.25 0.0285444744990429
4.75 0.0285444744990429
4.75 0
};
\addplot [red]
table {%
4.75 0.003439805323
5.25 0.00348139805323
};
\addplot [blue, mark=+, mark size=1, mark options={solid}, only marks]
table {%
5 0.0723684210526316
5 0.072992700729927
5 0.0739130434782609
5 0.0741797432239658
5 0.0743801652892562
5 0.0744125326370757
5 0.0746268656716418
5 0.0749279538904899
5 0.0761494252873563
5 0.0780730897009967
5 0.0782361308677098
5 0.0786516853932584
5 0.0790960451977401
5 0.0791476407914764
5 0.0794701986754967
5 0.0798319327731092
5 0.0802568218298555
5 0.0810092961487384
5 0.0845697329376855
5 0.0861812778603269
5 0.0868263473053892
5 0.0868794326241135
5 0.0870748299319728
5 0.0872675250357654
5 0.0874613003095975
5 0.0893416927899686
5 0.0893760539629005
5 0.0894187779433681
5 0.0906344410876133
5 0.0910290237467019
5 0.0913461538461538
5 0.0915395284327323
5 0.0916149068322981
5 0.0918544194107452
5 0.0919003115264797
5 0.0921787709497207
5 0.0923076923076923
5 0.0926470588235294
5 0.0986745213549337
5 0.0997229916897507
5 0.100502512562814
5 0.101361573373676
5 0.101404056162246
5 0.102564102564103
5 0.105785123966942
5 0.112328767123288
5 0.117241379310345
5 0.117355371900826
5 0.120200333889816
5 0.129518072289157
5 0.131326949384405
5 0.136701337295691
5 0.145161290322581
5 0.145772594752187
5 0.154605263157895
};
\addplot [blue, dashed]
table {%
6 0
6 0
};
\addplot [blue, dashed]
table {%
6 0.0307017543859649
6 0.0766666666666667
};
\addplot [black]
table {%
5.875 0.0766666666666667
6.125 0.0766666666666667
};
\addplot [black]
table {%
5.875 0
6.125 0
};
\addplot [blue]
table {%
5.75 0
6.25 0
6.25 0.0307017543859649
5.75 0.0307017543859649
5.75 0
};
\addplot [red]
table {%
5.75 0.00328139805323456
6.25 0.00328139805323456
};
\addplot [blue, mark=+, mark size=1, mark options={solid}, only marks]
table {%
6 0.0792602377807133
6 0.0795947901591896
6 0.0798376184032476
6 0.0802469135802469
6 0.0811188811188811
6 0.0819672131147541
6 0.0822981366459627
6 0.0840108401084011
6 0.0841121495327103
6 0.0842787682333874
6 0.0848
6 0.0864055299539171
6 0.08731241473397
6 0.0876369327073552
6 0.087667161961367
6 0.0902140672782875
6 0.0906593406593407
6 0.0920502092050209
6 0.0941176470588235
6 0.0942562592047128
6 0.0948553054662379
6 0.0953757225433526
6 0.096128170894526
6 0.0973724884080371
6 0.0980066445182724
6 0.0984375
6 0.0989956958393113
6 0.100470957613815
6 0.101246105919003
6 0.102106969205835
6 0.103633916554509
6 0.104815864022663
6 0.107458912768647
6 0.10797342192691
6 0.108433734939759
6 0.10857908847185
6 0.110378912685338
6 0.117117117117117
6 0.117962466487936
6 0.12015503875969
6 0.128664495114007
6 0.1328
6 0.134943181818182
6 0.135172413793103
6 0.148902821316614
6 0.148993288590604
6 0.150862068965517
6 0.1632
6 0.184100418410042
};
\end{axis}
\end{tikzpicture}
	\caption{{\em Relay adversary}: the distribution of compromise rates} 
	\label{fig:relay-circ}
\end{figure}

For the relay-level adversary, we randomly mark 10\% of our guard bandwidth and 10\% of our exit bandwidth as malicious guards and malicious exit relays in the network, then we simulate CAR, {\em vanilla}, and using three to five circuits with {\em RTT Only}. We run 10 simulations, where the malicious guards and exits change in each run. 10 simulations for each case is reasonable considering that we have 52 exit relays, 49 guards, and simulations taking 11 hours. 
Figure~\ref{fig:relay-circ} shows the distribution of stream compromised rates for clients. The median of compromised streams is almost the same for {\em vanilla}, CAR, and {\em RTT Only}  with {\em no change} in the number of circuits because the clients build almost the same number of circuits. As the number of circuits increases the median of compromised streams rate increases due to the increase in circuits created by the clients. When the number of circuits increases, the chance of creating a circuit which has a malicious relays on its guard and exit position increases. As we see, {\em RTT Only} with 5 circuits has the highest compromised streams.

\subsection{Network Level Adversary}\label{network-level-circ-selection}

In the network-level adversary model, we assume that the adversary controls an Autonomous System (AS). If the entry traffic (traffic between the client and guard) and exit traffic (traffic between the exit and server) traverse over a common AS, that AS can apply traffic correlation attack and link the client to her destinations.
\begin{figure}[t!]
	\centering
	\input{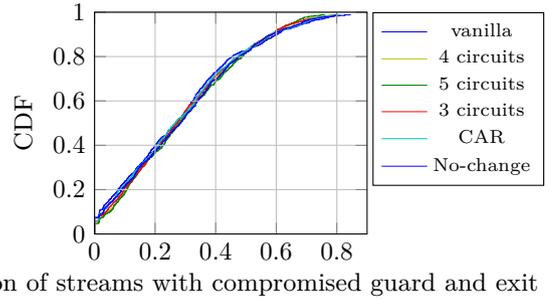}
	\caption{{\em Network adversary}: CDF of compromise rates} 
	\label{fig:network-circ}
\end{figure}

To analyze the security of circuit selection strategies, we used our simulation results from the relay-level adversary for  CAR, {\em vanilla}, and each circuit numbers in {\em RTT Only} in Shadow. Because we did not add any relays to the network in evaluating the relay-level adversary model and we only marked existing relays as malicious we can re-use their results in this selection. For each circuit selection approach, we extract all the generated streams by clients. The simulations generated approximately 730,000 streams for each circuit selection approach, or about 700 streams per client. For each stream, we used the algorithm proposed by Qiu and Gao~\cite{Qiu05aspath} to infer the AS paths between clients and guards and between exits and servers. This algorithm exploits known paths from BGP tables to improve the inferred paths. In measuring the compromise rates, we consider the possibility of asymmetric traffic correlation attack that can happen between the {\em data}  path and {\em ack} path, which is one of the RAPTOR attacks proposed by Sun et al. \cite{raptor}.

Figure \ref{fig:network-circ} shows the cumulative distribution of compromise rates for each strategy. As we see, when we increase the number of circuits from 3 to 5, the median compromise rate increases from 27.2\% to 28.1\% which are 0.7\% to 4\% worse than CAR. CAR performs 1\% better than {\em vanilla}. These results show that using RTT and increasing the number of circuits even up to five circuits have a very limited effect on the security in the network level adversary. Figure \ref{fig:network-circ} shows the cumulative distribution of compromise rates for each strategy.

\begin{figure*}[t]
        \centering
        \begin{subfigure}[ht]{0.32\textwidth}
                \includegraphics[trim={1.3cm 0.8cm 0.5cm 0.3cm},clip,width=0.95\textwidth]{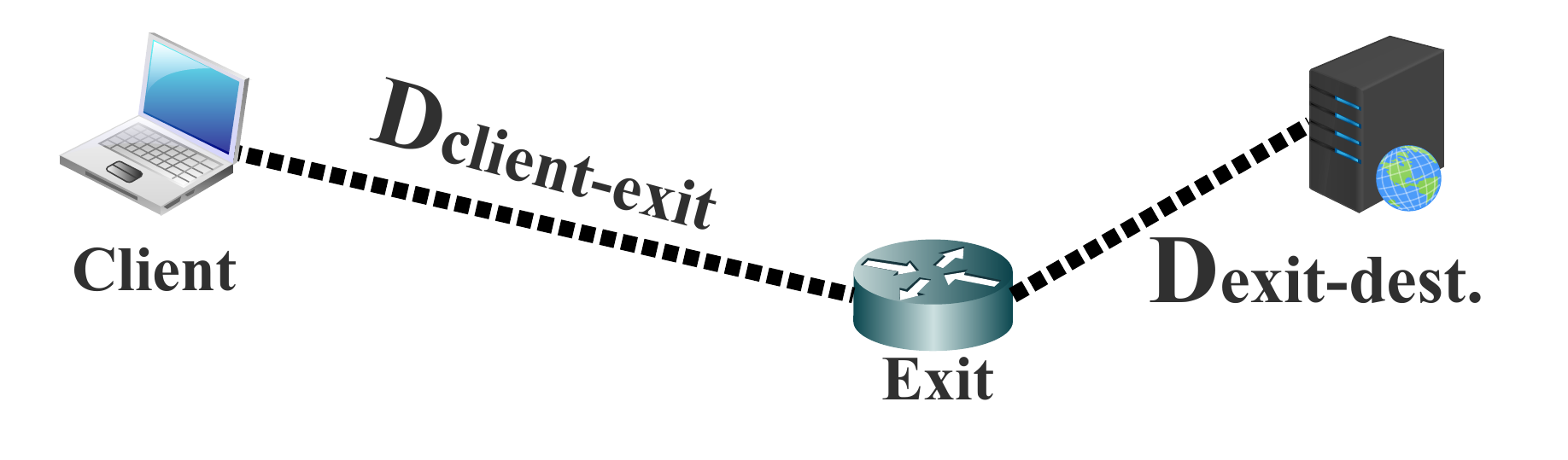}
                \caption{Exit}
                \label{fig:exit}
        \end{subfigure}\hspace{0.2cm}
        \begin{subfigure}[ht]{0.32\textwidth}
                \includegraphics[trim={1.2cm 0.8cm 0.5cm 0.3cm},clip,width=0.95\textwidth]{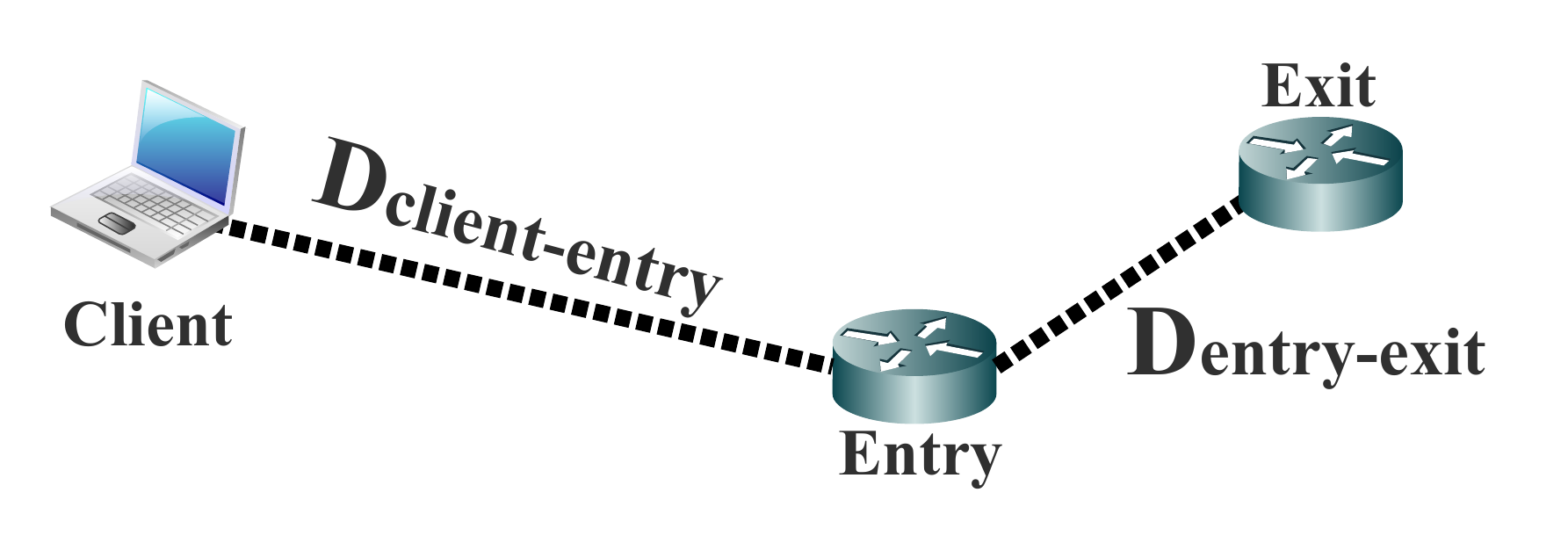}
                \caption{Entry}
                \label{fig:entry}
        \end{subfigure}\hspace{0.2cm}
        \begin{subfigure}[ht]{0.32\textwidth}
                \includegraphics[trim={1.3cm 0.8cm 0.5cm 0.3cm},clip,width=0.95\textwidth]{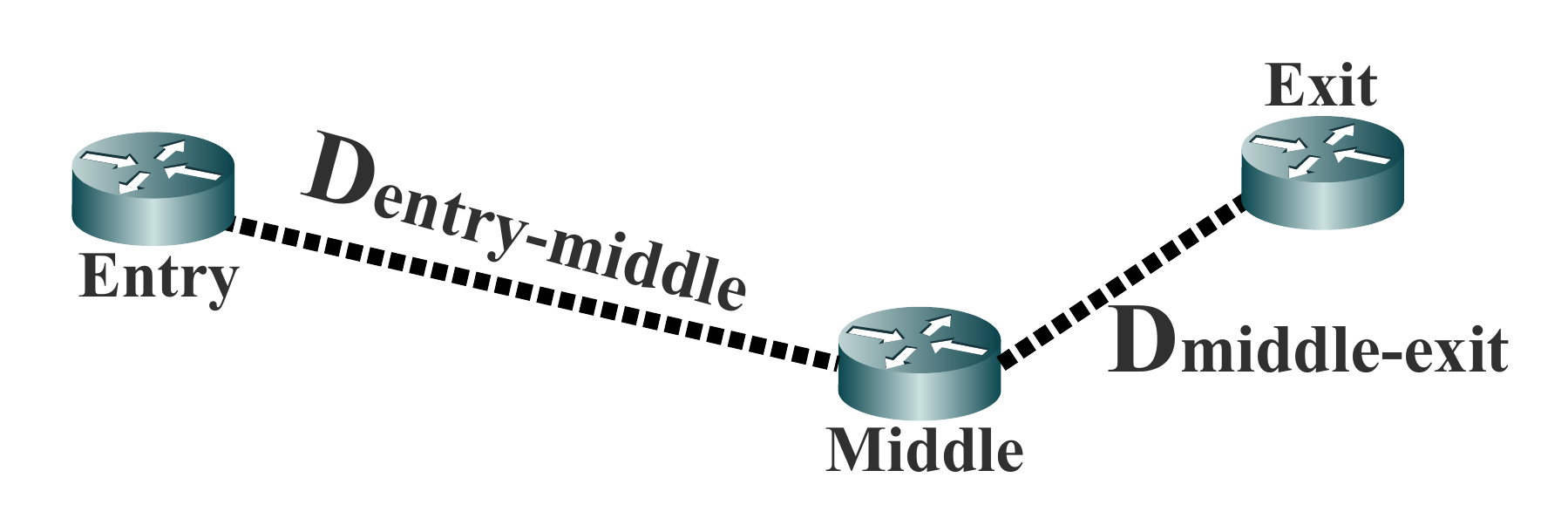}
                \caption{Middle}
                \label{fig:middle}
        \end{subfigure}
        \caption{Distance for relays for each position.}\label{fig:distance}
\end{figure*}

\section{Relay Selection}\label{relay-selection}


In this section, we describe a method for path selection in which we assign weights to relays based on a combination of bandwidth and geographical distances. This approach extends the idea of Tor path selection, which uses weights based on various factors in probabilistic relay selection. The goal of the combined weighting approach is to build circuits that still have high bandwidth relays, ensuring load balancing and good throughput, but also relatively shorter paths between the client and her destinations. 

\subsection{Weight Function}\label{weights2}
In a large and growing network like Tor, which consists of around 7000 nodes as of August 2016, examining all possible paths to find ones with these characteristics would be expensive. Clustering relays into geographic areas, as in LASTor~\cite{lastor}, can lead to uneven distribution of bandwidth between clusters. Instead, we approach the problem much like Tor's current algorithm by selecting one relay at a time, starting with the exit node, then the entry node, and finally the middle node. This greedy approach may miss the optimal path, for some definition of optimal, but our design aims to select from a wide range of paths with good performance and to avoid poorly performing paths. A broader selection of paths should help us maintain anonymity.

We calculate the weight of each relay using following function:
\begin{equation}
w= \alpha \times  w_B + (1 - \alpha) \times w_{D} \nonumber
\end{equation}
\noindent where $w_B$ is a measure of the relay's bandwidth and $w_{D}$ is a measure of distance. In this function, $\alpha$ is parameter that we can use to tune the share of bandwidth and distance in the weights. As $\alpha$ increases, the importance of bandwidth to the weight increases, and as $\alpha$ decreases, the importance of distance to the weight increases. $w_B$ and $w_{D}$ for relay \textit{i} are computed as follows:
\begin{equation}\label{eqn:wbw}
{w_B}_{i} = \frac{B_{i}}{B_{max}},  \;\;
{w_{D}}_{i} =1 - \frac{D_{i}}{D_{max}} \nonumber
\end{equation}


\noindent Here, $B_i$ is the relay's weighted bandwidth, and $B_{max}$ is the maximum weighted bandwidth among all relays. Tor assigns weights for each position in the circuit, and these weights bias the relay selection for circuits to distribute more load to higher-bandwidth relays. 
$D_{i}$ is a distance that is computed differently depending on the selected relay's role in the circuit as exit, middle, or entry. The maximum value of $D_{i}$ over all relays is $D_{max}$. We subtract the ratio from 1 so as to weight short distances more than long ones.
Note that both $w_B$ and $w_D$ will be between 0 and 1.

Figure~\ref{fig:distance} shows how we compute the distance for relays for each position in the circuit. We seek to minimize total distance from the client to the destination by minimizing the intermediate pairs of distances that are added by each relay in the sequence used by Tor: exit, entry, middle. Intuitively, selecting one of these relays with a large distance to its neighbors extends the path away from a straight line between client and destination, which would theoretically be the ideal path.

Since we use geographic locations, we compute $D$ using the great-circle distance between two points on a sphere from their longitudes and latitudes. In choosing the circuit's exit node, we compute $D$ for all relays as follows:
\begin{equation}
D_{exit} = (1 - \lambda) \times D_{client-exit} + \lambda \times D_{exit-dest} \nonumber
\end{equation}
For the circuit's entry node:
\begin{equation}
D_{entry} = \lambda \times D_{client-entry} + (1 -  \lambda) \times D_{entry-exit} \nonumber
\end{equation}
And for the circuit's middle node
\begin{equation}
D_{middle} = D_{entry-middle} + D_{middle-exit} \nonumber
\end{equation}

\noindent $D_{max}$ in $w_{D}$ is the maximum computed \textit{D} among the set of relays for each position (exit, entry, or middle). The use of $D_{max}$ and $B_{max}$ ensure that $w_D$ and $w_B$ both range between 0 and 1 for more straightforward calculations. 

\begin{figure}
        \centering
        \begin{subfigure}[t]{0.5\textwidth}                \includegraphics[width=0.95\textwidth, trim={0.8cm 0.8cm 0.6cm 0.3cm},clip]{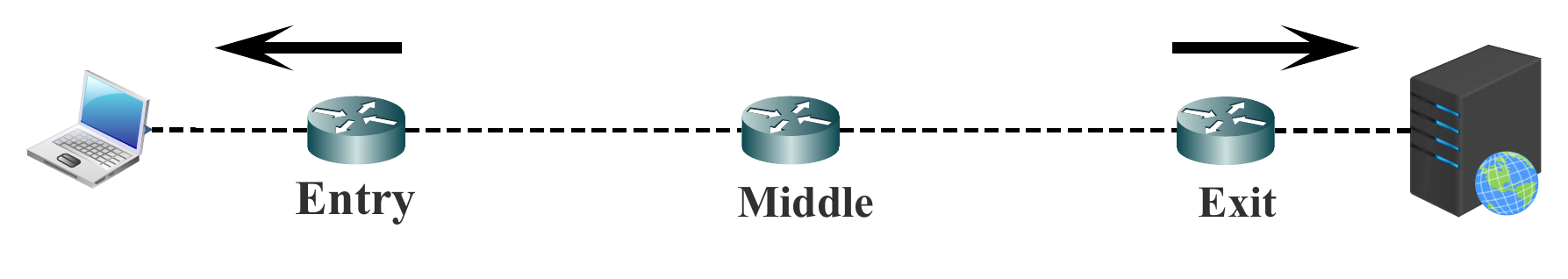}
                \caption{Large value of $\lambda$}
                \label{fig:lambdahigh}
        \end{subfigure}\\
        \begin{subfigure}[t]{0.5\textwidth}
        \includegraphics[width=0.95\textwidth,trim={0.8cm 0.8cm 0.6cm 0},clip]{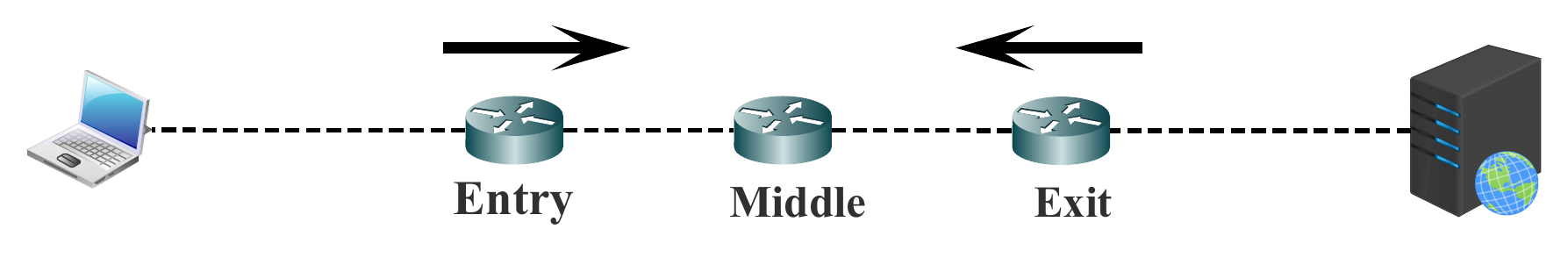}
                \caption{Small value of $\lambda$}
                \label{fig:lambdalow}
        \end{subfigure}
        \caption{The effect of $\lambda$.}\label{fig:lambdavaries}
\end{figure}

$\lambda$ is a tuning parameter that enables us to change the share of the distance between different nodes on the path. As shown in Figure~\ref{fig:lambdavaries}, as $\lambda$ goes up, the entry nodes and exit nodes move toward the source and destination, respectively, and a large portion of the path between the source and destination is covered by inter-relay connections. In this case, the selected guards are close to the clients which can decrease the threat of network level adversaries. In particular, it causes the AS paths between clients and guards to be shorter and involve fewer ASes, thereby decreasing the chance that a common AS appears on both sides of the traffic, i.e. between the client and guard and between the exit and server. We will evaluate the effect of nearby guards in Sections~\ref{as-adv-relay-selection} and Appendix ~\ref{nearby}. As $\lambda$ decreases, the path's inter-relay portion shrinks. 


\begin{figure}
  \centering
 \includegraphics[width=0.37\textwidth,height=3.7cm,trim={0 0 0 0.7cm},clip]{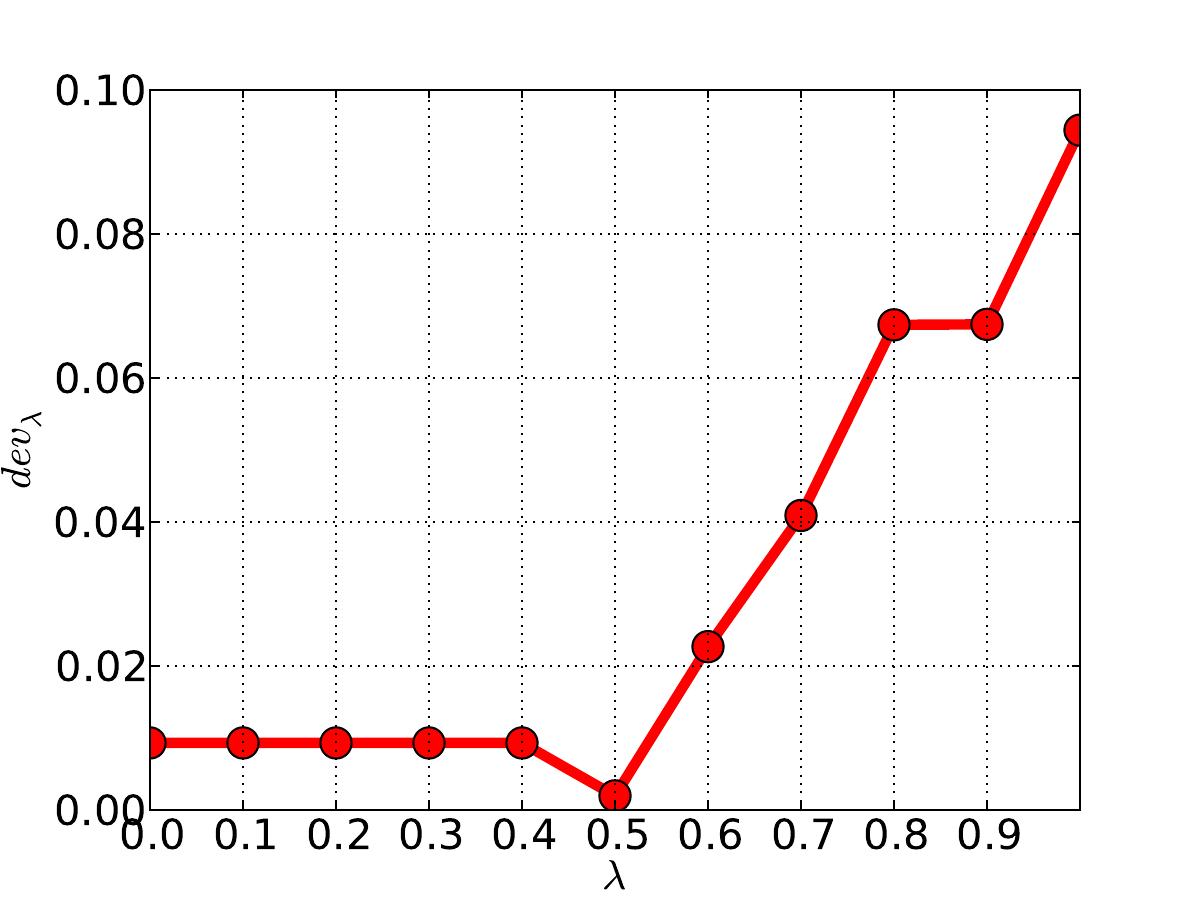}
  \caption{Average added distance over the shortest path for different
    values of $\lambda$.}
  \label{fig:lambda}
\end{figure}

To evaluate how close the short paths selected by our algorithm are to optimal, we used a scaled-down Tor network, with 147 exit nodes, 700 middle nodes, and 170 entry nodes, a user located in the central US, and 100 destinations from the Alexa Top 100 websites~\cite{alexa}. We found short paths between the user and all destinations with our method for different values of $\lambda$. To measure the average distance between our method from the actual shortest path to each of these 100 destinations ($d$), we used the mean absolute percentage deviation (MAPD): 
\begin{equation}
dev_{\lambda} =\frac{1}{100} \sum_{d=1}^{d=100}\frac{|L_{\lambda d} - L_{d}|}{L_d}
\nonumber
\end{equation}
where $L_{\lambda d}$ is the length of the shortest path found by our method for a given value of $\lambda$, and $L_d$ is the length of the shortest path to destination $d$. Figure~\ref{fig:lambda} shows the average deviations $dev_{\lambda}$ for different values of $\lambda$, and we see that $\lambda = 0.5$ has the lowest deviations at just 0.19\% longer on average. This indicates that our greedy algorithm produces short paths close to the optimal ones.

In Appendix ~\ref{nearby}, We use our relay selection approach to implement the nearby guard proposal, a defense mechanism proposed by Sun et al. ~\cite{raptor},  on TorPS, the Tor path simulator~\cite{johnson:ccs13}.  

\subsection{Preemptively built circuits}\label{circuits}
When computing distances, we need to know the location of the final destination. The destination addresses can be either an IP address or a DNS hostname. If the address is IP, we can find the destination's location by using IP geolocation databases (we use Maxmind \cite{maxmind} in this paper) and find the shortest path to these destinations. But most of time, the addresses are DNS hostnames. To find the location, we first need to perform DNS resolution to get the IP address. In the DNS resolution process, the party performing DNS lookup returns the closest content provider or replica sever to itself, which in Tor's case means the closest ones to the exit nodes. 

In Tor, however, the client saves time by having a number of circuits already built and available for new connections. Building a new circuit takes time for numerous protocol messages and public-key cryptographic operations. Wacek et al. showed that LASTor, which builds new circuits once the destination location is discovered, suffers from significant added delay due to these delays~\cite{empirical}.

\begin{figure}[t]
  \centering
  \includegraphics[width=0.35\textwidth]{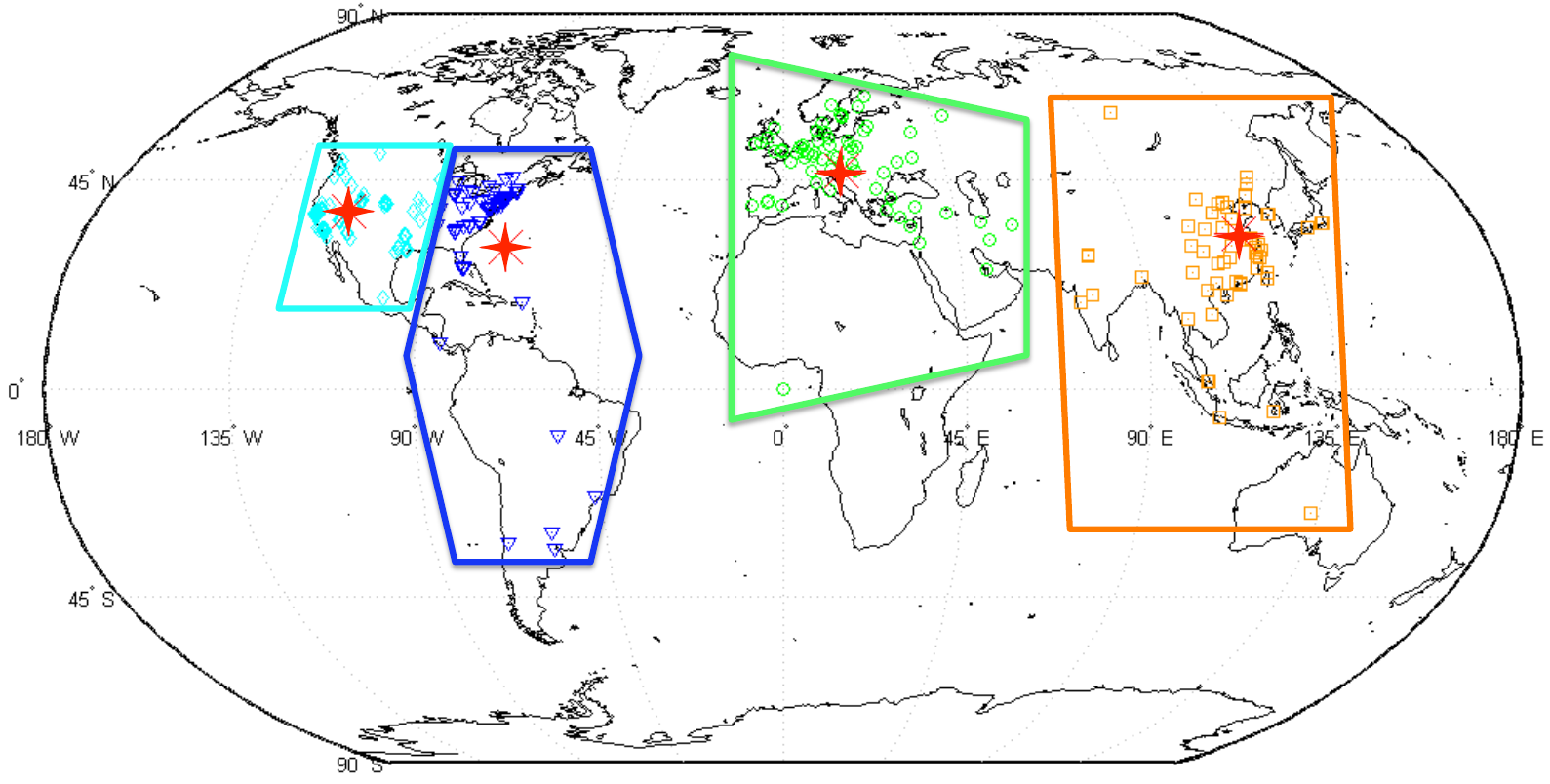}
  \caption{The location of the existing servers in Alexa Top 1000 websites. Red stars show the cluster centroids.} 
  \label{fig:cluster}
\end{figure}
To solve this issue, we build circuits in advance that shorten the path between the source and the most popular destinations online. To identify popular destinations, we use the Alexa Top 1000 websites~\citep{alexa}. Because some of these sites use CDNs or replica servers, we visited them from different places in thew world. In particular, we selected 13 planet-lab nodes based on Tor users' statistics~\cite{Tor-users}. From each node, we visited all of the 1000 sites, got the address of all fetchable elements in their front pages, and resolved their addresses if needed. For example, from our planet-lab node on the US west coast, we got 46,306 hostname addresses, leading to 2894 unique IP addresses from 346 unique locations. We clustered the obtained locations into four clusters using the k-means algorithm (we got better performance for four rather than with three or five clusters), and we use the centroids of these clusters as a target destination. Figure~\ref{fig:cluster} shows all obtained locations and our cluster centroids. From these four target destinations, we mark the one closest to the client as the default destination, i.e. when we have no other information, we assume that the users is more likely to visit sites located closer to her. We have the client build circuits in advance of their use with short paths to these four target destinations. We check our circuit list once per second to ensure that we have at least one circuit to each of these destinations and ensure that a new connection can be handled as quickly as possible.
 
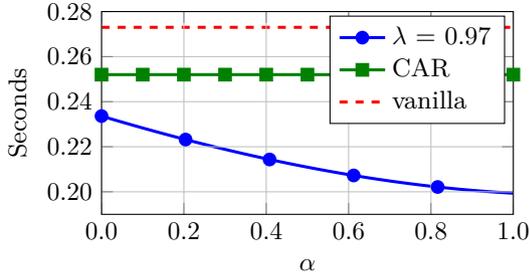
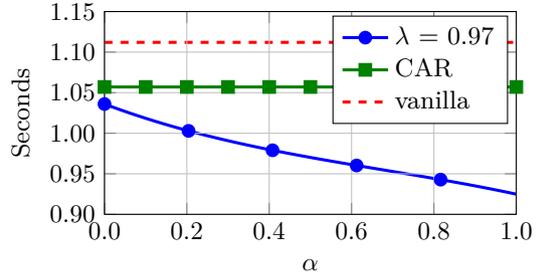
\begin{figure*}[t]
        \centering
        \begin{subfigure}[ht]{0.47\linewidth}
               \centering
%
%
%
\begin{tikzpicture}
\centering
\begin{axis}[
ylabel={Seconds},
xlabel={$\alpha$},
xmin=0, xmax=1,
ymin=0.19, ymax=0.28,
axis on top,
width=\figurewidth,
height=0.85\figureheight,
xmajorgrids,
ymajorgrids,
xtick={0,0.2,0.4,0.6,0.8,1},
xticklabels={$0.0$,$0.2$,$0.4$,$0.6$,$0.8$,$1.0$},
ytick={0.18,0.2,0.22,0.24,0.26,0.28,0.3},
yticklabels={,$0.20$,$0.22$,$0.24$,$0.26$,$0.28$,},
legend entries={{$\lambda$ = 0.97},{CAR},{vanilla}},
legend cell align={left}
]
\addplot [blue,mark=*,mark repeat={10}, line width = 0.4mm]
table {%
0 0.233606182161397
0.0204081632653061 0.232500378345135
0.0408163265306122 0.231408388220105
0.0612244897959184 0.230330413078623
0.0816326530612245 0.229266654213005
0.102040816326531 0.228217312915567
0.122448979591837 0.227182590478623
0.142857142857143 0.22616268819449
0.163265306122449 0.225157807355483
0.183673469387755 0.224168149253917
0.204081632653061 0.223193915182108
0.224489795918367 0.222235306432372
0.244897959183673 0.221292524297025
0.26530612244898 0.220365770068381
0.285714285714286 0.219455245038756
0.306122448979592 0.218561150500466
0.326530612244898 0.217683687745827
0.346938775510204 0.216823058067153
0.36734693877551 0.215979462756761
0.387755102040816 0.215153103106966
0.408163265306122 0.214344180410084
0.428571428571429 0.21355289595843
0.448979591836735 0.21277945104432
0.469387755102041 0.212024046960069
0.489795918367347 0.211286884997993
0.510204081632653 0.210568166450407
0.530612244897959 0.209868092609627
0.551020408163265 0.209186864767969
0.571428571428572 0.208524684217748
0.591836734693878 0.20788175225128
0.612244897959184 0.207258270160879
0.63265306122449 0.206654439238863
0.653061224489796 0.206070460777546
0.673469387755102 0.205506536069243
0.693877551020408 0.204962866406271
0.714285714285714 0.204439653080945
0.73469387755102 0.203937097385581
0.755102040816326 0.203455400612493
0.775510204081633 0.202994764053999
0.795918367346939 0.202555389002412
0.816326530612245 0.202137476750049
0.836734693877551 0.201741228589226
0.857142857142857 0.201366845812257
0.877551020408163 0.201014529711459
0.897959183673469 0.200684481579147
0.918367346938776 0.200376902707636
0.938775510204082 0.200091994389242
0.959183673469388 0.199829957916281
0.979591836734694 0.199590994581068
1 0.199375305675919
};
\addplot [very thick, green!50.0!black, mark=square*, line width = 0.4mm]
table {%
0 0.252
0.1 0.252
0.2 0.252
0.3 0.252
0.4 0.252
0.5 0.252
0.6 0.252
0.7 0.252
0.8 0.252
0.9 0.252
1.00 0.252
};
\addplot [very thick, red, dashed, line width = 0.4mm]
table {%
0 0.273
100 0.273
};
\end{axis}
\end{tikzpicture}
                \caption{TTFB}
                \label{fig:alpha-web-ttfb}
        \end{subfigure}
        \begin{subfigure}[ht]{0.47\linewidth}
               \centering
%
%
%
\begin{tikzpicture}
\centering
\begin{axis}[
ylabel={Seconds},
xlabel={$\alpha$},
xmin=0, xmax=1,
ymin=0.9, ymax=1.15,
axis on top,
width=\figurewidth,
height=0.85\figureheight,
xmajorgrids,
ymajorgrids,
xtick={0,0.2,0.4,0.6,0.8,1},
xticklabels={$0.0$,$0.2$,$0.4$,$0.6$,$0.8$,$1.0$},
ytick={0.9,0.95,1,1.05,1.1,1.15},
yticklabels={$0.90$,$0.95$,$1.00$,$1.05$,$1.10$,$1.15$},
legend entries={{$\lambda$ = 0.97},{CAR},{vanilla}},
legend cell align={left}
]
\addplot [blue,mark=*,mark repeat={10}, line width = 0.4mm]
table {%
0 1.03597402290961
0.0204081632653061 1.03214213176266
0.0408163265306122 1.0284373068275
0.0612244897959184 1.02485559828942
0.0816326530612245 1.02139305633373
0.102040816326531 1.01804573114571
0.122448979591837 1.01480967291067
0.142857142857143 1.0116809318139
0.163265306122449 1.0086555580407
0.183673469387755 1.00572960177636
0.204081632653061 1.00289911320619
0.224489795918367 1.00016014251548
0.244897959183673 0.997508739889535
0.26530612244898 0.99494095551364
0.285714285714286 0.992452839573098
0.306122448979592 0.990040442253206
0.326530612244898 0.987699813739262
0.346938775510204 0.985427004216562
0.36734693877551 0.983218063870403
0.387755102040816 0.981069042886084
0.408163265306122 0.9789759914489
0.428571428571429 0.97693495974415
0.448979591836735 0.974941997957131
0.469387755102041 0.972993156273139
0.489795918367347 0.971084484877473
0.510204081632653 0.969212033955428
0.530612244897959 0.967371853692303
0.551020408163265 0.965559994273395
0.571428571428572 0.963772505884
0.591836734693878 0.962005438709417
0.612244897959184 0.960254842934941
0.63265306122449 0.958516768745872
0.653061224489796 0.956787266327505
0.673469387755102 0.955062385865138
0.693877551020408 0.953338177544068
0.714285714285714 0.951610691549593
0.73469387755102 0.949875978067009
0.755102040816326 0.948130087281614
0.775510204081633 0.946369069378705
0.795918367346939 0.944588974543579
0.816326530612245 0.942785852961533
0.836734693877551 0.940955754817866
0.857142857142857 0.939094730297873
0.877551020408163 0.937198829586852
0.897959183673469 0.935264102870101
0.918367346938776 0.933286600332916
0.938775510204082 0.931262372160595
0.959183673469388 0.929187468538435
0.979591836734694 0.927057939651733
1 0.924869835685787
};
\addplot [line width = 0.4mm, green!50.0!black, mark=square*]
table {%
0 1.057
0.1 1.057
0.2 1.057
0.3 1.057
0.4 1.057
0.5 1.057
0.6 1.057
0.7 1.057
0.8 1.057
0.9 1.057
1.00 1.057
};
\addplot [line width = 0.4mm,red, dashed]
table {%
0 1.112
100 1.112
};
\end{axis}
\end{tikzpicture}
                \caption{TTLB}
                \label{fig:alpha-web-ttlb}
        \end{subfigure}
        \caption{Median TTFB and TTLB for web clients}\label{fig:alpha-web}
\end{figure*}

\subsection{Guard Selection}\label{Guards}
In Tor, each client selects and uses one guard consistently for a period of nine to ten months. This means that for selecting the path, the guard relay is already selected, before the exit. 
We thus cannot use the exit's location to help pick the entry and must modify our algorithm. In selecting the guard, instead of using the exit node's location, we use the closest of the four target destinations to the client to compute $D_{entry}$. 
$D_{entry}$ will be computed as:
\begin{equation}
D_{entry} = \lambda \times D_{client-entry} + (1 -  \lambda) \times D_{entry-target} \nonumber
\end{equation}
This can reveal some information about the client's location, e.g. through fingerprinting attacks that identify the guard from the exit~\cite{tissec-latency-leak}. Since we only have four popular destinations, however, the anonymity sets for clients' location will be quite large. We evaluate the security of our design in Section~\ref{security_analysis-relay-selection}. Further performance improvement can be achieved by using $D_{guard-exit}$ instead of $D_{client_exit}$ in computing $D_exit$ because we have already selected the guard relay and know its location. This helps us to not have long circuits in case the guard is relatively far from the client.

During this research, we found a bug in the Tor source code that was causing Tor clients to choose guards from all the available relays, not the ones with the guard flag. This harmed both anonymity and performance. The bug was reported to the Tor project\footnote{\url{https://trac.torproject.org/projects/tor/ticket/17772}} and was fixed in Tor version 2.7.6. 
We use the corrected code in all of our simulations.

\subsection{Attaching streams to the circuits}\label{attach-relay-selection}

In Section~\ref{attaching-streams}, we introduced and evaluated different strategies of attaching streams to the circuits, and we found that using {\em RTT Only} offers the best performance. Our relay selection mechanism tries to find the short paths between the client and the destination. Therefore, we use {\em RTT then length} to pick the fast and short circuits in the rest of the work. 

An alert reader may wonder why we do not use {\em RTT} to construct circuits given its strong performance in selecting completed circuits. We note that RTT is not available before constructing the circuit, and so it cannot be used when picking relays. Sherr et al. used estimated latencies for their relay selection algorithm~\cite{Sherr09}, but Wacek et al. found that CAR was more effective than this technique~\cite{modeling}. In our approach, we seek to build more circuits, most of which should have reasonable performance, and then use RTT measurements to select the current best circuit from this set. This provides the benefits of using RTT with less risk of not having any high-performing circuits.

\section{Performance Evaluation}\label{experiment-relay-selection}
In this section, we evaluate the performance of our proposed relay selection method compared with Tor and congestion-aware routing (CAR), the current state of the art~\cite{empirical}. 
In all the experiments in this section, we use Shadow with the same network configuration as used in Section~\ref{experiment-circ-selection}. We evaluate the performance of our method as $\alpha$ and $\lambda$ vary.

\paragraphX{The effect of {\large $\alpha$}.}\label{alpha}
For evaluating different values of $\alpha$, we set $\lambda=0.97$ and vary $\alpha$ from 0.0 to 1.0, we have chosen $\lambda$ high to stretch circuits between the clients and destinations. 
Figure~\ref{fig:alpha-web} shows the median of web clients' TTLB and TTFB with respect to $\alpha$, where we plot only the median of the results for readability. Changing $\alpha$ from 0.0 to 1.0 yields 7\% to 24\% improvement in web clients' TTFB compared to CAR, and 2\% to 12\% improvement in web clients' TTLB compared to CAR. 
Relay bandwidth is more important in improving performance, which matches findings from Wacek et al.~\cite{empirical}. We will explore the trade-offs in security in Section~\ref{security_analysis-relay-selection}.

\paragraphX{The effect of {\large $\lambda$}.}\label{lambda}
Parameter $\lambda$ controls the elasticity of the path. As $\lambda$ increases, the circuit stretches out, with guards moving toward the clients and exits moving toward the destinations. To evaluate the effect of $\lambda$, we set $\alpha=0.0$. We find that performance changes less than 2\% as $\lambda$ varies, and it is best for $\lambda$ of 0.4-0.5.

\section{Security analysis}\label{security_analysis-relay-selection}

We now examine the impact of our path selection strategies on anonymity using three models. First, we study broad system-wide measures of anonymity. We then examine path compromise rates in the presence of AS adversary. Finally, we study a set of targeted attacks in the relay level adversary.

\subsection{System-wide Security Metrics}\label{gini-entropy}
We simulated the proposed path-selection strategies in a 2127-relay model of the Tor network, built by sampling approximately one-third of the nodes of each type (exit, entry, middle) from a descriptor file from December 2015. In our simulations, 200 clients are placed according to statistics about users from the Tor metrics portal, and each client constructs 27,000 paths. In our location-based approaches, the clients select paths using the four target destinations as described in Section~\ref{circuits}. We created in total 5.4 million paths for different values of $\alpha$, with $\lambda$ = 0.97. 

\begin{table}
 \centering
\begin{tabular}{r @{\hskip 0.3cm} | c c}
  Strategy  &  Gini Coef.   &   Entropy \\
  \hline
{\em vanilla}&0.725&9.702\\ 
$\alpha = 1.0$&0.724&9.713\\ 
$\alpha = 0.9$&0.590&10.326\\ 
$\alpha = 0.8$&0.490&10.609\\ 
$\alpha = 0.7$&0.444&10.735\\ 
$\alpha = 0.6$&0.401&10.802\\ 
$\alpha = 0.5$&0.370&10.872\\ 
$\alpha = 0.15$&0.335&10.933\\
$\alpha = 0.1$&0.338&10.944\\ 
$\alpha = 0.0$&0.341&10.945 \\ 
\end{tabular}
	\caption{{\bf Relay adversary.} System-wide security results.} 
    \vskip -0.5cm
\end{table}\label{table:security_table}

\begin{figure*}[t]
        \centering
        \begin{subfigure}[ht]{0.45\textwidth}
%
%
%
\begin{tikzpicture}
\definecolor{color0}{rgb}{0.75,0.75,0};
\begin{axis}[
xlabel={relay selection},
ylabel={Median},
xtick={0,20,40,60,80,100},
xticklabels={0,0.2,0.4, 0.6,0.8, 1.0},
xmin=0, xmax=100,
ymin=0.24, ymax=0.28,
axis on top,
width=\figurewidth,
height=0.85\figureheight,
xmajorgrids,
ymajorgrids,
legend style={at={(0.03,0.97)}, anchor=north west},
legend entries={{$\lambda = 0.97$},{CAR},{vanilla}},
legend cell align={left}
]
\addplot [black, mark=*,mark repeat={10}, line width = 0.4mm]
table {%
0 0.250621529820906
2.04081632653061 0.250184260723248
4.08163265306122 0.2497497325452
6.12244897959184 0.249319877306094
8.16326530612245 0.248896627025263
10.2040816326531 0.248481913722042
12.2448979591837 0.248077669415762
14.2857142857143 0.247685826125757
16.3265306122449 0.247308315871361
18.3673469387755 0.246947070671905
20.4081632653061 0.246604022546724
22.4489795918367 0.24628110351515
24.4897959183673 0.245980245596516
26.530612244898 0.245703380810156
28.5714285714286 0.245452441175403
30.6122448979592 0.245229358711589
32.6530612244898 0.245036065438049
34.6938775510204 0.244874493374114
36.734693877551 0.244746574539118
38.7755102040816 0.244654240952394
40.8163265306122 0.244599424633275
42.8571428571429 0.244584057601094
44.8979591836735 0.244610071875185
46.9387755102041 0.24467939947488
48.9795918367347 0.244793972419513
51.0204081632653 0.244955722728417
53.0612244897959 0.245166582420924
55.1020408163265 0.245428483516368
57.1428571428571 0.245743358034081
59.1836734693878 0.246113137993398
61.2244897959184 0.246539755413651
63.265306122449 0.247025142314173
65.3061224489796 0.247571230714297
67.3469387755102 0.248179952633357
69.3877551020408 0.248853240090685
71.4285714285714 0.249593025105615
73.469387755102 0.250401239697479
75.5102040816327 0.251279815885611
77.5510204081633 0.252230685689344
79.5918367346939 0.253255781128011
81.6326530612245 0.254357034220944
83.6734693877551 0.255536376987478
85.7142857142857 0.256795741446945
87.7551020408163 0.258137059618678
89.7959183673469 0.259562263522011
91.8367346938776 0.261073285176276
93.8775510204082 0.262672056600807
95.9183673469388 0.264360509814936
97.9591836734694 0.266140576837997
100 0.268014189689323
};
\addplot [ very thick, color0, dashed, line width = 0.4mm]
table {%
0 0.266666666666667
10 0.266666666666667
20 0.266666666666667
30 0.266666666666667
40 0.266666666666667
50 0.266666666666667
60 0.266666666666667
70 0.266666666666667
80 0.266666666666667
90 0.266666666666667
100 0.266666666666667
};
\addplot [ very thick, green!50.0!black,  mark=square*, line width = 0.4mm]
table {%
0 0.26984126984127
10 0.26984126984127
20 0.26984126984127
30 0.26984126984127
40 0.26984126984127
50 0.26984126984127
60 0.26984126984127
70 0.26984126984127
80 0.26984126984127
90 0.26984126984127
100 0.26984126984127
};
\end{axis}
\end{tikzpicture}
                \caption{As $\alpha$ varies}
                \label{fig:alpha-relay-AS}
        \end{subfigure}
        \begin{subfigure}[ht]{0.45\textwidth}
%
%
%
\begin{tikzpicture}
\definecolor{color0}{rgb}{0.75,0.75,0};
\begin{axis}[
xlabel={relay selection},
ylabel={Median},
xtick={0,20,40,60,80,100},
xticklabels={0,0.2,0.4, 0.6, 0.8, 1.0},
xmin=0, xmax=100,
ymin=0.24, ymax=0.28,
axis on top,
width=\figurewidth,
height=0.85\figureheight,
xmajorgrids,
ymajorgrids,
legend style={at={(0.6,0.97)}, anchor=north west},
legend entries={{$\alpha = 0$},{CAR},{vanilla}},
legend cell align={left}
]
\addplot [black,mark=*,mark repeat={10}, line width = 0.4mm]
table {%
0 0.257585277944131
2.04081632653061 0.258435819695709
4.08163265306122 0.259174918368602
6.12244897959184 0.259806551625905
8.16326530612245 0.260334697130718
10.2040816326531 0.260763332546137
12.2448979591837 0.261096435535258
14.2857142857143 0.26133798376118
16.3265306122449 0.261491954887
18.3673469387755 0.261562326575814
20.4081632653061 0.26155307649072
22.4489795918367 0.261468182294814
24.4897959183673 0.261311621651196
26.530612244898 0.26108737222296
28.5714285714286 0.260799411673205
30.6122448979592 0.260451717665028
32.6530612244898 0.260048267861526
34.6938775510204 0.259593039925796
36.734693877551 0.259090011520935
38.7755102040816 0.258543160310041
40.8163265306122 0.257956463956211
42.8571428571429 0.257333900122541
44.8979591836735 0.25667944647213
46.9387755102041 0.255997080668073
48.9795918367347 0.255290780373469
51.0204081632653 0.254564523251415
53.0612244897959 0.253822286965008
55.1020408163265 0.253068049177344
57.1428571428571 0.252305787551522
59.1836734693878 0.251539479750638
61.2244897959184 0.250773103437789
63.265306122449 0.250010636276074
65.3061224489796 0.249256055928588
67.3469387755102 0.248513340058429
69.3877551020408 0.247786466328695
71.4285714285714 0.247079412402481
73.469387755102 0.246396155942887
75.5102040816327 0.245740674613008
77.5510204081633 0.245116946075943
79.5918367346939 0.244528947994787
81.6326530612245 0.243980658032639
83.6734693877551 0.243476053852596
85.7142857142857 0.243019113117754
87.7551020408163 0.242613813491211
89.7959183673469 0.242264132636065
91.8367346938776 0.241974048215411
93.8775510204082 0.241747537892348
95.9183673469388 0.241588579329973
97.9591836734694 0.241501150191383
100 0.241489228139674
};
\addplot [ very thick, color0, dashed, line width = 0.4mm]
table {%
0 0.266666666666667
100 0.266666666666667
};
\addplot [ very thick,green!50.0!black, mark=square*, line width = 0.4mm]
table {%
0 0.26984126984127
10 0.26984126984127
20 0.26984126984127
30 0.26984126984127
40 0.26984126984127
50 0.26984126984127
60 0.26984126984127
70 0.26984126984127
80 0.26984126984127
90 0.26984126984127
100 0.26984126984127
};
\end{axis}
\end{tikzpicture}
                \caption{As $\lambda$ varies}
                \label{fig:lambda-relay-AS}
        \end{subfigure}
        \caption{{\bf AS Adversary.} The median stream compromised rate as $\alpha$ and $\lambda$ vary}\label{fig:relay-as}
\end{figure*}
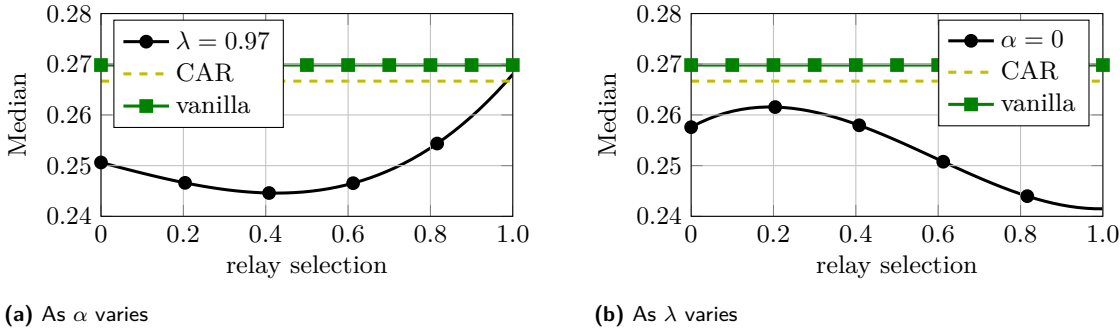

To measure anonymity, we focus on end-to-end traffic confirmation attacks in which the adversary controls both the exit and entry relays in a circuit. We measured the Gini coefficient and Shannon entropy of the exit-entry combinations occurring on selected circuits. The Gini coefficient is a measure of the equality of relay selection, where 0 represents pure equality (i.e. each relay is selected uniformly at random) and 1 represents a state of complete inequality (i.e. a given relay is always selected)~\cite{tune_up,empirical}. Table \ref{table:security_table} shows the results for Gini coefficient and entropy.

In our static path selection method, we cannot produce equivalent results for CAR, since circuits dynamically change in their method. Wacek et al. report that CAR has a lower (i.e., better) Gini coefficient than Tor but slightly lower (i.e., worse) entropy~\cite{empirical}. As shown in Table~\ref{table:security_table}, as $\alpha$ decreases, the Gini coefficient decreases and the entropy increases. As we expect, for $\alpha$ = 1.0, which selects the relays only based on bandwidth, the results are nearly the same as {\em vanilla}. The most dramatic difference in anonymity occurs for higher values of $\alpha$, e.g. from $\alpha=1.0$ and $\alpha=0.8$, where the Gini coefficient drops from 0.724 to 0.490 and entropy rises 0.9 bits. 

Since $\alpha$ means increasing the share of distance in the selection weights, we see that emphasizing distance in the weights improves the system-wide security metrics. On the other hand, according to Figure ~\ref{fig:alpha-web}, small values of $\alpha$ offer lower performance. 
We thus face a trade-off between security and performance, where decreasing $\alpha$ improves security but offers less performance benefits. Values of $\alpha$ between 0.8 and 0.5 offer both good security and performance, with a Gini coefficient of between 0.370-0.490 and almost 20\% improvement in TTFB compared to CAR.   

\subsection{AS Adversary}\label{as-adv-relay-selection}
In this section, we evaluate the security of our approach in the presence of AS-level adversaries. The same as our security analysis in Section~\ref{network-level-circ-selection}, we use Shadow with the same configuration as previous sections and carry out simulations for different $\alpha$ and $\lambda$ values. 

For evaluating the effect of $\alpha$, we fix $\lambda=0.97$ and vary $\alpha$ from 0.0 to 1.0. For each value of $\alpha$, we extract all the generated streams along with their attached circuits in the simulation. For all the streams, we find the AS paths between the clients and guards (guards and clients) and between the exits and the destinations (destinations and exits) using the algorithm proposed by Qiu and Gao~\cite{Qiu05aspath}. 
We consider the possibility of an asymmetric traffic correlation attack that can happen between the {\em data} path and {\em ack} path. Figure~\ref{fig:alpha-relay-AS} shows the median stream compromise rates.
As we see, by increasing $\alpha$ from 0, the compromise rate starts decreasing until $\alpha$  reaches around 0.5, where we have the minimum compromise rate. After $\alpha$ = 0.5, the compromise rate again increases until we have the worst case at $\alpha$ = 1.0, which is close to {\em vanilla}'s and CAR's compromise rates. The compromise rate in CAR is slightly better than {\em vanilla}. 

To evaluate the effect of $\lambda$ on the security, we set $\alpha=0.0$ and vary $\lambda$ from 0 to 1.0. Figure~\ref{fig:lambda-relay-AS} shows the median compromise rate as $\lambda$ varies. As we expect, when $\lambda$ increases, the compromise rate generally decreases. This results from stretching the circuits closer to the
communication end points, which in turn reduces the chance of common ASes appearing on both the entry and exit sides traffic.

\subsection{Targeted Attacks in Relay-Level Adversary}
To further explore how path selection strategies perform against attacks in the relay level model, we now examine four types of attacks in the network and measured how often adversaries can compromise a circuit. We assume that a circuit is compromised if both the exit and entry nodes are controlled by the adversary. 

In the targeted attacks, we consider a high-bandwidth adversary that owns a few high-bandwidth relays such that its  bandwidth is a considerable fraction of the network's total bandwidth. In particular, we start with our 2127-relay model of the Tor network as described in Section~\ref{gini-entropy}. We select a random bandwidth in the range [20 MiB/s, 220 MiB/s], where 220 MiB/s is the maximum bandwidth in our model network. A malicious relay with this bandwidth is added to the Tor network, where the location of the malicious relay is based on our attack strategies. This process is repeated until the target attacker bandwidth for this {\em run} of the experiment is reached. For each of 200 runs at each bandwidth setting, we place one client into the network using a location based on Tor metrics data and have it pick 9,000 paths for each of our tested strategies. We use the four popular destinations as described in Section~\ref{circuits} for our strategies. In our evaluations, we consider four different attack strategies as follows:


\begin{enumerate}
\item \textit{Targeted Clients}: In this attack strategy, in each run that we add the client and malicious relays, all the malicious guards are located in the exact location of the client, just as if the adversary could run all of his guards in the client's room. The malicious exit relays are randomly placed in locations based on the geographical distribution of Tor relays.

\begin{figure*}
  \centering
 \includegraphics[scale=0.3]{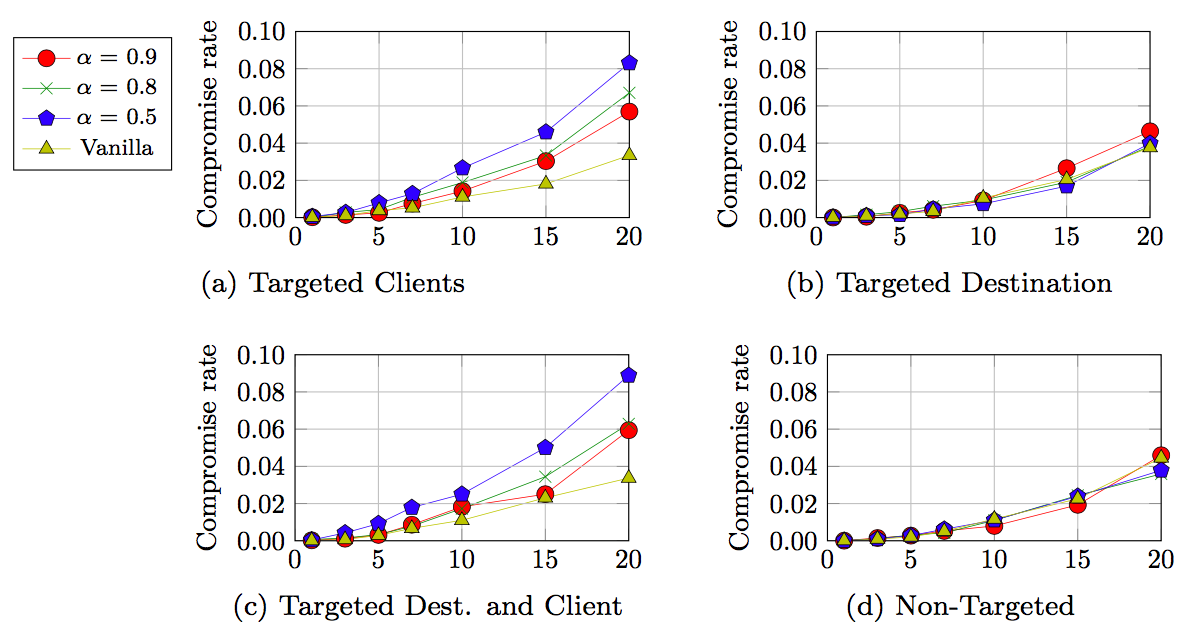}
  \caption{{\bf Targeted Attacks:} Fraction of circuits compromised. X-axis shows the percentage of exit and guard bandwidth controlled by the attacker.}\label{fig:targeted-attacks}
\end{figure*}

\item \textit{Targeted Destination}: In this attack strategy, in each run that we add the malicious relays, all the malicious exits are located in the exact location of the one of the randomly selected popular destinations, just as if  the adversary could run all of its exit relays in the same server room. The malicious guard relays are randomly placed in locations based on Tor relays geographical distribution.

\item \textit{Targeted Destination and Client}: In this attack strategy, in each run that we add the client and the malicious relays, all the malicious exits are located in the exact location of the one of the randomly selected popular destinations, and all the malicious guards are located in the exact location of the client.

\item \textit{Non-targeted}: In this attack strategy, in each run that we add  the malicious relays, all the malicious relays are randomly placed in locations based on the geographical distribution of Tor relays.
\end{enumerate}
Figure~\ref{fig:targeted-attacks} shows the fraction of compromised paths with respect to the percentage of total bandwidth controlled by the adversary's relays for {\em vanilla},$\alpha$  = 0.5,$\alpha$  = 0.8, and $\alpha$  = 0.9.
As shown in Figure~\ref{fig:targeted-attacks}, the compromise rate for {\em Targeted Destination} is almost the same as compromise rates in {\em vanilla}. For {\em Targeted Client} and for {\em Targeted Destination and Client}, the compromise rates for $\alpha$ = 0.5,0.8, and 0.9 are worse than {\em vanilla}, and as the adversary's bandwidth fraction increases, the gap between them and {\em vanilla} increases. For both {\em Targeted Client} and for {\em Targeted Destination and Client}, $\alpha$ = 0.8 has almost the same compromise rate as $\alpha$ = 0.9 but better compromise rate than $\alpha$ = 0.5.  For {\em Non-targeted} we observed the same compromise rate as {\em vanilla} for $\alpha$ = 0.5,0.8, and 0.9 because in this attack malicious relays are randomly located in the network and all relay selection methods pick the malicious relays with the same probability.  

We also note that this is a trade-off with the modest, but wide-spread, security benefits of using $\alpha=0.8$ on Gini coefficient, entropy, and compromise rates compared with $\alpha=1.0$, for example. Greater emphasis on bandwidth leads to better performance and more resilience to targeted attacks, while greater emphasis on distance leads to more diffuse spreading of load on the network. 


\section{Discussion}\label{discuss}
In this section, we discuss the implications of our findings and the scope for future research.

\paragraphX{Circuit Selection.}
We evaluated the impact of different number of pre-built circuits on Tor performance, and found that having at least three pre-built circuits ready results in a significant improvement compared to {\em vanilla} Tor. Preparing more than three circuits, however does not provide much additional benefit and may also add more load on the network. Our circuit selection mechanisms also kill unused circuits after five minutes, which raises the rate of exploring for better circuits.


\paragraphX{Relay Selection.}
In relay selection, combined weighting seems to provide a trade-off of performance and anonymity. As the weights emphasize on the bandwidth, $\alpha$ is high, combined weighting provides higher performance. On the other hand, higher values of $\alpha$ could not provide diverse paths. As $\alpha$ goes down and the weights are inclined toward the distances, the performance improvement decreases, but the created circuits are more diverse. Low values $\alpha$ suffer from a greater chance of targeted relay-level attacks. Overall, we think that combined weighting with $\alpha=0.8$ seems to provide the best trade-off of performance and anonymity. The best value of $\alpha$ may vary with network configuration, bandwidth distribution, geographical dispersion of relays, and the client's location. We will examine setting $\alpha$ more carefully in future work. 

\paragraphX{Nearby guards.}
Our evaluations showed that the proposed defense, picking guards close to the clients, does not effect an AS-level adversary. The AS path between the clients and guards is not highly correlated with the geographical distance between them. The AS path length between the guards and clients depend on the clients' networks, guards' networks, and their ASes relationships with other ASes. Moreover, the clients and guards are not uniformly distributed on the globe and on the network. For example, a single AS, AS16276, is contributing more than 170 guards to the Tor network, which is 16\% of all the guards in January 2015. The other issue is guard rotation, as currently Tor clients change their guard after 9 to 10 months. In our 10-month TorPS simulations, the median number of guard changes for clients was five times, with a minimum of two times and a maximum of 29 times. Thus, even if the client is secure due to the short AS path, after guard rotation, she may pick a guard that has a long AS path length and get compromised. We evaluate the nearby guards in Appendix~\ref{nearby}.

\bibliographystyle{abbrv}
\bibliography{refs}
\appendix

\section{Nearby Guards}\label{nearby}
Tor is known to be vulnerable to adversaries observing the entry and exit traffic, since a simple traffic correlation attack~\cite{levine04timing,ccs2013-usersrouted,raptor} can link the user to the destination (an \textit{end-to-end attack}). If an adversary sits on the first and the last hop of the path, the adversary can see both sides of the traffic and carry out the traffic correlation attack. 
The adversary may perform this attack by controlling components of the network, like Autonomous Systems (ASes) or Internet Exchange Points (IXPs). As a defense against such AS-level adversaries, Sun et al. suggest picking guards close to the clients~\cite{raptor}. This results in having fewer ASes in the path on the entry side of the traffic and can decrease the chance that an adversary can observe both ends of the traffic. 

\subsection{Attacker Model.}
We evaluate the security of nearby guards in terms of both relay-level and network-level adversaries.
\paragraphX{Relay-Level Adversary Model.}
We examine the security of nearby guard in the presence of relay-level adversaries. We use TorPS in this evaluation and consider the case that the attacker targets a specific user. We analyze two relay-level adversaries, high bandwidth and low bandwidth adversaries. In low bandwidth adversary we consider that the attacker adds one low bandwidth guard to the network and places it in the target's location, and in high bandwidth adversary, the attacker adds one high bandwidth guard in the target's location. These two evaluations can show how much bandwidth and distance matter in compromising streams. 
\paragraphX{Network-Level Adversary Model.}
To analyze the security of nearby guards in the network level, we simulate the proposed algorithm in TorPS to get the paths. We follow the same methodology as our network-level adversary in circuit selection and relay selection in obtaining the AS paths and measuring compromise rates.
\subsection{Network Model.}
To evaluate the consequences on security of using guard nodes close to the client, 
we use TorPS, 
a tor path simulator that uses realistic models to mimic users' web browsing behavior, and historical data to model the network. 
TorPS 
builds the circuits using Tor's path selection code and real consensuses and server descriptors.
In our TorPS simulation, we consider 30 client ASes and picked the client ASes from top client ASes found by Edman et al.~\cite{ASawareness} and Tor Metrics~\cite{Tor-users}. We ran the simulations in two time periods, for one month of Tor descriptor data each, from February 2015 to March 2015, with 500 clients in each of 30 client ASes (15,000 clients total), and for 10 months of Tor descriptor data, from February 2015 to November 2015, with 100 clients in each of 30 client ASes (3,000 clients total).

\subsection{Implementation of nearby guards}\label{close-guards}

Malicious ASes can perform traffic correlation attacks passively over the traffic passing through them, or they could use one of RAPTOR attacks~\cite{raptor} to actively hijack the traffic and put themselves in the paths between the guard and the client. Sun et al. proposed as a defense against these attacks is to pick guards close to the clients~\cite{raptor}, since short paths mean fewer ASes and thus less chance of malicious ASes being on the path between the client and the guard. This approach has never been evaluated. On the other hand, if clients select nearby guards, it means that a relay-level adversary can run some guards close to a targeted client and increase his chance to be selected. This suggests a trade-off between the two attacks. In this section, we examine this trade-off in detail.

We partially evaluated the effect of picking close guards in Section \ref{as-adv-relay-selection}, and we showed that a large value of $\lambda$ reduces the compromise rate in the presence of an AS-level adversary. Since changing $\lambda$ also affects the path to the destination, we now isolate the study to picking guards close to the client.


\begin{figure}[t]
   \centering
                \input{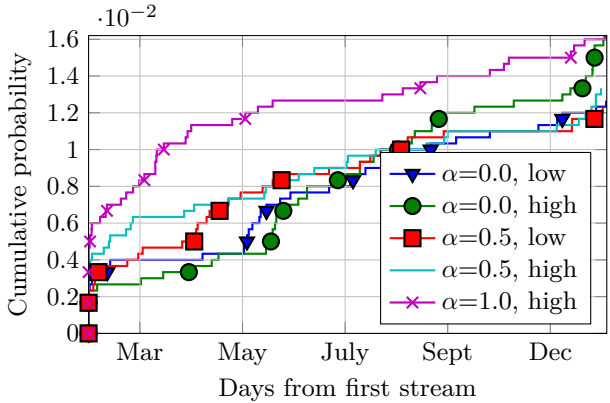}
                \caption{Time to first compromise in 10-month simulations}
                \label{fig:relay-10-month}
\end{figure}

To examine the impact of close guards, we use the TorPS simulator~\cite{johnson:ccs13} to generate streams and build circuits. TorPS is a path simulator that uses real Tor data and realistic models to mimic Tor path selection behavior. We used the {\em typical} user model in the TorPS configuration, which consists of Gmail, Google Chat, Google Calendar, Docs, Facebook, and web search activity. We consider 30 client ASes in our simulation and pick these ASes in a way that covers both top client ASes as identified by Edamn et al.~\cite{edman:ccs09} and the geographical distribution of Tor users~\cite{torusers}. 
In our model, we place 100 users on each AS (3,000 users total).
We modified the TorPS path selection module to implement the selection of nearby guards. We change the weight of candidate relays for the guard position in the relay selection module according to our weighting function from Section~\ref{weights2} ($w= \alpha \times  w_B + (1 - \alpha) \times w_{D}$). The only difference is in $D_{entry}$, we remove parameter $\lambda$ and redefine the distance as $D_{entry}=D_{client-entry}$, the distance between client and guard. In other words, we ignore the distance to the destination because we want to only evaluate the proximity of guards to the clients not destinations. Note that we only change the weights for guard selection; the selection process for other positions, exit and middle, remains the same as {\em vanilla}.
\begin{figure}
  \centering
 \includegraphics[width=70mm,height=43mm]{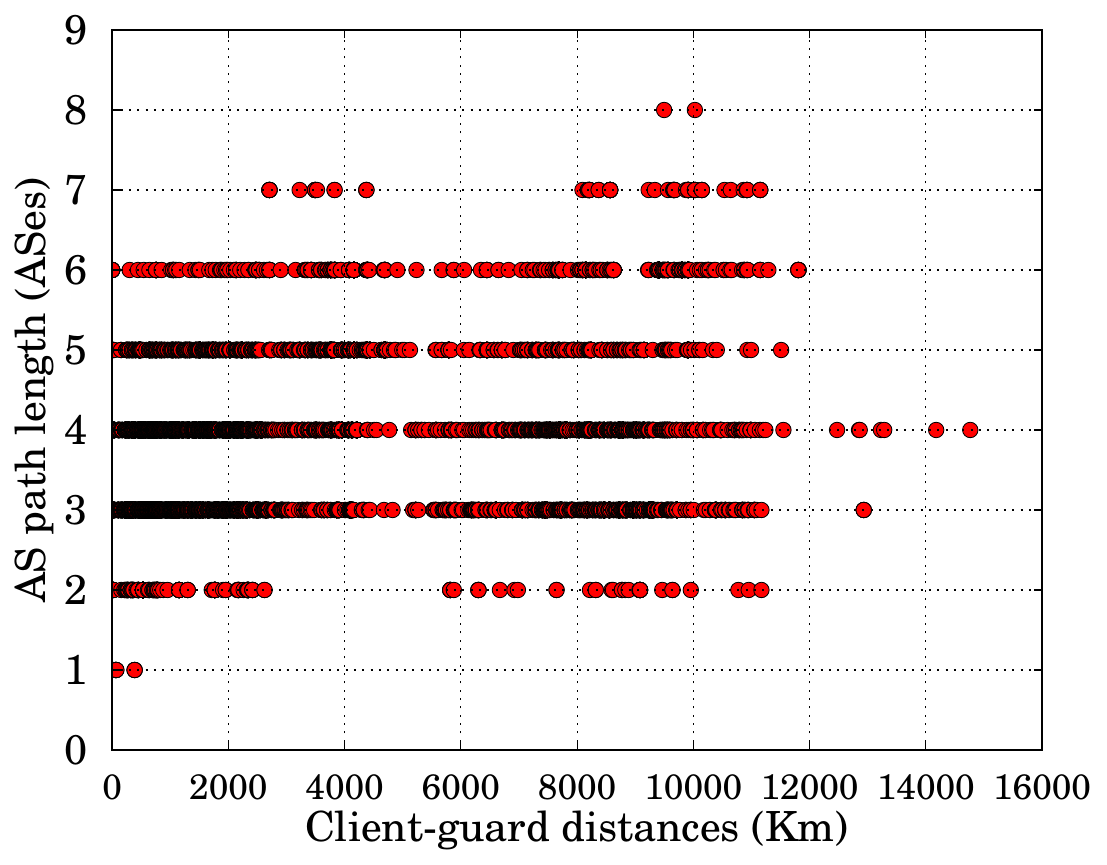}
  \caption{AS path length}\vskip -0.6cm
  \label{fig:as-path-length}
\end{figure}
\subsection{Relay-level adversary}\label{relay-close-guard}
In evaluating the threat of relay-level adversaries, we use the methodology of Johnson et al.~\cite{johnson:ccs13} and considered an adversary running a  guard relay.\footnote{We only analyze  the guard compromise, the case the clients select the malicious guards, not full compromise that the clients pick both malicious guard and malicious exit. We assume that if the client picks a malicious guard, some of her streams passing through that guard will be compromised.}
When $\alpha$ is low, the guard's bandwidth does not have significant role in the selection weight. On the other hand, for high values of $\alpha$, the guard's bandwidth plays a notable role in the weights and guard selection. Given these facts, we considered a {\em high bandwidth} and a {low bandwidth} adversary in our evaluations. The {\em high bandwidth} adversary injects a guard relay with 55 MBps bandwidth, which is in the top 5\% of guard bandwidths as of February 2015. The {\em low bandwidth} adversary adds a guard relay with 2MBps bandwidth, which is in the bottom 10\% of guard bandwidth and the minimum bandwidth for getting the guard flag. In both cases, for each client we placed the malicious guard in the client's precise location (i.e. $D_{entry}=D_{client-entry} = 0$).


Figure \ref{fig:relay-10-month} shows the cumulative probability of the time to the first stream get compromised for different cases. As expected, both {\em low bandwidth} and {\em high bandwidth} adversaries have the same time to first compromise for $\alpha$ = 0.0 because the share of bandwidth is zero in weights and the distance only matters. In {\em high bandwidth} adversary, when $\alpha$ increases, time to first compromise decreases, as shown $\alpha$ = 1.0 has the worst time to first compromised guard. In the {\em low bandwidth} adversary,when $\alpha$ is low the malicious guard's weight is higher and has more chance to be picked. As shown in the figure, the time to first compromised guard is almost the same for $\alpha$ equal to 0.5 and 0.0 because $W_D$  is dominated the the weights. For $\alpha$ = 1.0 in {\em low bandwidth} adversary, none of clients picked the malicious guard during the 10-month simulation.

\begin{figure}
  \centering
%
%
%
\begin{tikzpicture}

\definecolor{color0}{rgb}{0.75,0.75,0}

\begin{axis}[
ylabel={Frac. of compromised streams},
xlabel={$\alpha$},
xmin=0, xmax=1.03,
ymin=0, ymax=0.007,
axis on top,
xmajorgrids,
ymajorgrids,
width=\figurewidth,
height=\figureheight,
xtick={0,0.2,0.4,0.6,0.8,1,1.2},
legend style={at={(0.05,0.97)}, anchor=north west},
legend cell align={left},
legend entries={{low BW adversary},{high BW adversary}}
]
\addplot [very thin, blue, dashed]
table {%
0 0.00368777649655309
0.0204081632653061 0.0036928203942789
0.0408163265306122 0.00369847840832343
0.0612244897959184 0.00370451620608954
0.0816326530612245 0.00371069945498009
0.102040816326531 0.00371679382239794
0.122448979591837 0.00372256497574597
0.142857142857143 0.00372777858242702
0.163265306122449 0.00373220030984397
0.183673469387755 0.00373559582539968
0.204081632653061 0.00373773079649701
0.224489795918367 0.00373837089053881
0.244897959183673 0.00373728177492797
0.26530612244898 0.00373422911706733
0.285714285714286 0.00372897858435977
0.306122448979592 0.00372129584420813
0.326530612244898 0.0037109465640153
0.346938775510204 0.00369769641118412
0.36734693877551 0.00368131105311747
0.387755102040816 0.0036615561572182
0.408163265306122 0.00363819739088918
0.428571428571429 0.00361100042153327
0.448979591836735 0.00357973091655333
0.469387755102041 0.00354415454335223
0.489795918367347 0.00350403696933283
0.510204081632653 0.00345914386189799
0.530612244897959 0.00340924088845058
0.551020408163265 0.00335409371639346
0.571428571428571 0.00329346801312948
0.591836734693878 0.00322712944606152
0.612244897959184 0.00315484368259243
0.63265306122449 0.00307637639012509
0.653061224489796 0.00299149323606234
0.673469387755102 0.00289995988780706
0.693877551020408 0.00280154201276211
0.714285714285714 0.00269600527833035
0.73469387755102 0.00258311535191464
0.755102040816326 0.00246263790091785
0.775510204081633 0.00233433859274284
0.795918367346939 0.00219798309479247
0.816326530612245 0.0020533370744696
0.836734693877551 0.0019001661991771
0.857142857142857 0.00173823613631783
0.877551020408163 0.00156731255329465
0.897959183673469 0.00138716111751044
0.918367346938775 0.00119754749636803
0.938775510204082 0.000998237357270313
0.959183673469388 0.000788996367620136
0.979591836734694 0.000569590194820367
1 0.000339784506273868
};
\addplot [very thin, color0]
table {%
0 0.003341999312744
0.0204081632653061 0.00350517718028589
0.0408163265306122 0.00364757104113819
0.0612244897959184 0.00377018391560782
0.0816326530612245 0.00387401882400172
0.102040816326531 0.0039600787866268
0.122448979591837 0.00402936682379
0.142857142857143 0.00408288595579823
0.163265306122449 0.00412163920295843
0.183673469387755 0.00414662958557751
0.204081632653061 0.0041588601239624
0.224489795918367 0.00415933383842004
0.244897959183673 0.00414905374925733
0.26530612244898 0.00412902287678122
0.285714285714286 0.00410024424129861
0.306122448979592 0.00406372086311645
0.326530612244898 0.00402045576254165
0.346938775510204 0.00397145195988114
0.36734693877551 0.00391771247544184
0.387755102040816 0.00386024032953068
0.408163265306122 0.00380003854245459
0.428571428571429 0.00373811013452049
0.448979591836735 0.0036754581260353
0.469387755102041 0.00361308553730595
0.489795918367347 0.00355199538863937
0.510204081632653 0.00349319070034248
0.530612244897959 0.0034376744927222
0.551020408163265 0.00338644978608547
0.571428571428571 0.0033405196007392
0.591836734693878 0.00330088695699032
0.612244897959184 0.00326855487514576
0.63265306122449 0.00324452637551244
0.653061224489796 0.00322980447839729
0.673469387755102 0.00322539220410722
0.693877551020408 0.00323229257294918
0.714285714285714 0.00325150860523008
0.73469387755102 0.00328404332125685
0.755102040816326 0.00333089974133641
0.775510204081633 0.00339308088577568
0.795918367346939 0.0034715897748816
0.816326530612245 0.00356742942896109
0.836734693877551 0.00368160286832107
0.857142857142857 0.00381511311326848
0.877551020408163 0.00396896318411022
0.897959183673469 0.00414415610115324
0.918367346938775 0.00434169488470445
0.938775510204082 0.00456258255507078
0.959183673469388 0.00480782213255915
0.979591836734694 0.00507841663747649
1 0.00537536909012973
};
\addplot [draw=blue, fill=blue, mark=*, only marks] table {%
0.0 0.00369983535968
0.1 0.0034413627153
0.2 0.00400426798379
0.3 0.00424526342452
0.4 0.00298537234043
0.5 0.00388851950355
0.6 0.00259816995947
0.7 0.00280923885512
0.8 0.00246481762918
0.9 0.0015781724924
1.0 0.000128147163121
};
\addplot [draw=color0, fill=color0, mark=*, only marks] table {%
0.0 0.00341204407295
0.1 0.0037490881459
0.2 0.00452246073961
0.3 0.00395989108409
0.4 0.00318940602837
0.5 0.00390459726444
0.6 0.00357762158055
0.7 0.00320981509625
0.8 0.00389344604863
0.9 0.00318722771023
1.0 0.00581942755826
};
\end{axis}

\end{tikzpicture}
  \caption{Frac. of compromise streams as $\alpha$ varies in one-month simulations}\vskip -0.6cm
  \label{fig:relay-one-month}
\end{figure}
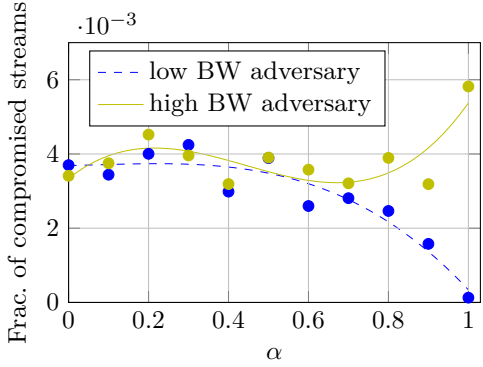

Figure~\ref{fig:relay-one-month} shows the fraction of compromised streams as $\alpha$ changes for one month simulation of both adversary models. As we see, the compromise rate is high for {\em high bandwidth} adversary  as $\alpha$ changes, because the malicious relay has high selection weight for both high and low values of $\alpha$. The compromise rate decreases for {\em low bandwidth} adversary  as $\alpha$ increases because the malicious relay's selection weight is low in high values of $\alpha$. 

\subsection{AS level adversary}\label{AS-close-guard}
In this section we assume that the adversary controls a single AS and applies the traffic correlation attack in order to de-anonymize users. Picking guards close to the clients is assumed to decrease the AS path between the client and the guard. As a result of this, the chance of appearing a common AS on the both entry and exit sides of the traffic seems to decrease. To evaluate this theory, we run TorPS modified with our guard selection scenario for different values of $\alpha$ for one months and and 10 months the same as the previous selection configurations. We find that the fraction of compromised streams is very high in all cases and almost completely independent of alpha. Clients will have a compromised stream within one hour of using Tor, and picking nearby guards does not help. 

An important reason for this is that the number of ASes is not significantly reduced by having shorter distances.
%
To see the relationship between the geographical distance (between guards and clients) and AS path lengths (the number of AS between guards and clients), we extracted all guards picked by our clients as $\alpha$ = 0.0 and computed their distance between guard-client pairs and their AS path lengths. Fig~\ref{fig:as-path-length} shows a scatter plot of guard-client distances and their AS path lengths. There is not a clear relationship or correlation between the distance and AS path lengths; for all short and long distances, the majority of paths lengths have three or four ASes. 

\section{Modified Tuning Function}\label{our-tune-up}

\begin{figure*}[ht]
        \centering
        \begin{subfigure}[ht]{0.3\textwidth}
                \includegraphics[width=50mm]{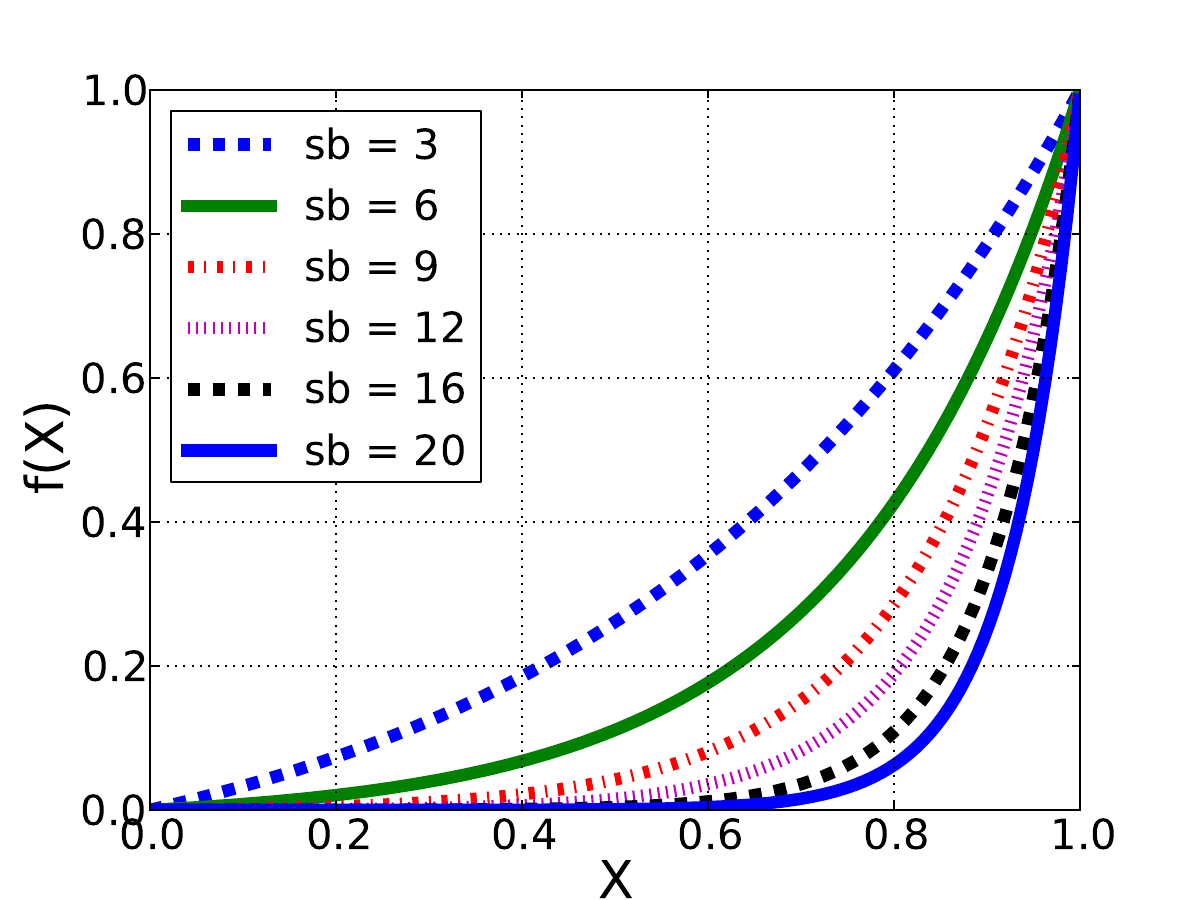}
                \caption{}
                \label{fig:tune_up}
        \end{subfigure}
        \begin{subfigure}[ht]{0.3\textwidth}
                \includegraphics[width=50mm]{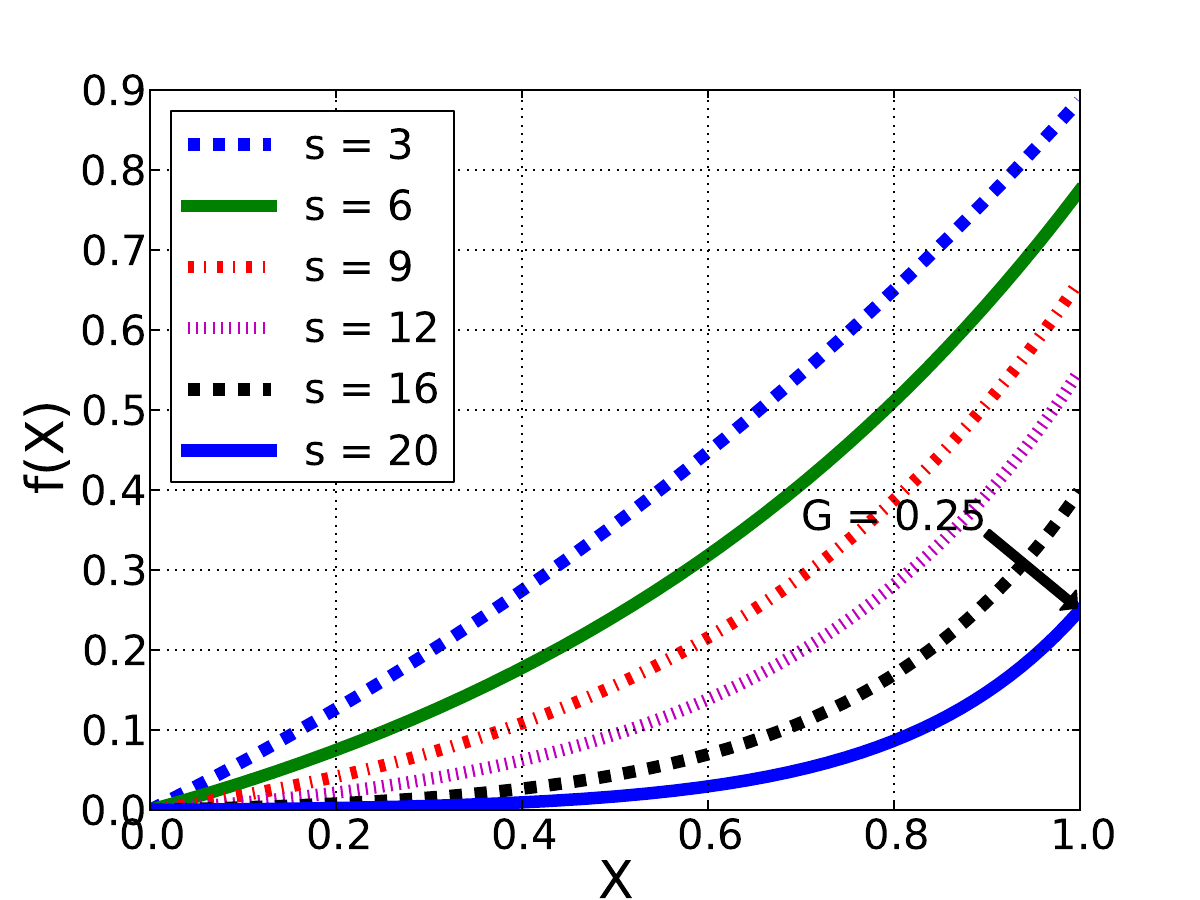}
                \caption{}
                \label{fig:s-varies}
        \end{subfigure}
        \begin{subfigure}[ht]{0.3\textwidth}
                \includegraphics[width=50mm]{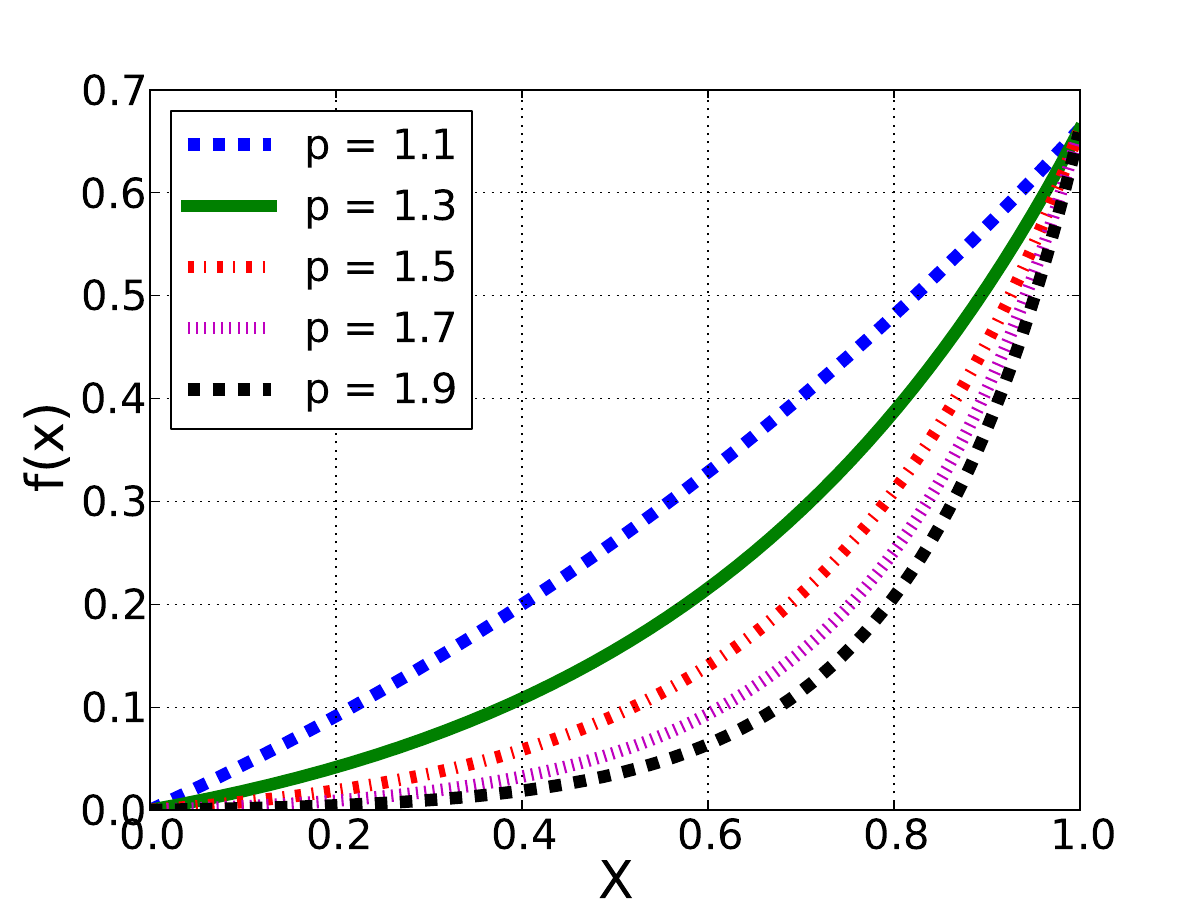}
                \caption{}
                \label{fig:p-varies}
        \end{subfigure}
        \caption{(a) Snader-Borisov family of functions; (b) Our tuning function as $s$ varies; (c) Our tuning function as $p$ varies. $G_{min} = 0.25$ and $S_{max} = 20$.}\label{fig:our_tune_up}
\end{figure*}

In this appendix, we describe our modification to the Snader-Borisov tuning function. Figure~\ref{fig:tune_up} shows the original Snader-Borisov family of functions for different values of \textit{s}. 

In this function, for a performance-concerned user there is still a chance to choose low bandwidth relays. When the number of performance-concerned users increase, high bandwidth relays would be over-utilized and become congested, which would result in these users actually experiencing worse performance. We modified this function to address these shortcomings as shown: 

\[f_{s}: [0 \ 1]\rightarrow [0 \ 1]
\]
\begin{equation}\label{eqn: selection function}
f_{s}(x) = \frac{1 - p^{sx}}{1 - p^{s}}\times \left(1 - \frac{1 - G_{min}}{S_{max}} \times s\right)
\end{equation}

In this function, $G_{min}$ ($G_{min}\in\mathbb [0 \ 1]$) determines the smallest pool of high weight relays that users are allowed to use in \textit{s} = $S_{max}$ ($S_{max}$ is maximum acceptable selection parameter), and \textit{p} lets us control the curvature of the graph. Figure \ref{fig:our_tune_up} shows the modified tuning function (the function presented by Snader et al.~\cite{tune_up} is a special case of our modified tuning function for $p = 2$ and $(G_{min} = 1$).

Controlling the curvature is useful when we need to balance the load on high weight relays if they are getting congested. Decreasing the curvature leads to decreasing the probability of choosing high weight relays, which it means we can control their load. This function also helps users to withdraw some portion of low weight relays depend on their selection parameter, but they cannot reach less than $(G_{min}\times 100)$ percent of high weight relays, which prevents them from choosing a small set of high weight relays. As shown in Figure \ref{fig:s-varies}, if we have $n$ relays, as \textit{s} increases from zero to $S_{max}$, the pool that users are allowed to select relays includes from all \textit{n} relays to $G_{min} \times n$ relays.

\section{Breaking Ties}\label{breaking-ties}

In Section~\ref{relay-selection}, we introduced a weighting function that mixes geographical distance and bandwidth to select relays for the circuit. Using a combined function to weight relays requires combining two rather different quantities---one being a measure of edges and another being a measure of the node---in a non-standard way. In this section, we propose a simple alternative: we first use one metric, bandwidth or distance, to narrow down the list of relays and then use the other metric to ``break the tie'' and pick among the smaller list. 

\subsection{Distance-first}\label{tie bandwidth}
In this approach, we first select a set of relays with low distance (where distance is defined depending on its position in the circuit) and then select a high-bandwidth relay from this set. As in combined weighting, we assign weights $w_B$ and $w_D$ to each relay as defined in Equation~\eqref{eqn:wbw}. To choose a relay for each position in the circuit, we first fill a bucket with $k$ relays, selected based on their distance weights $w_D$. In particular, we use our relay selection function (see Section~\ref{relay-selection}) with weights $w_D$ to pick $k$ relays from among all relays flagged for the given position in the circuit, i.e. exit, entry, or middle. Note that this is a probabilistic selection, not a deterministic one, i.e. we do not take the $k$ closest relays. Then among the \textit{k} relays in the bucket, we again use our relay selection function with weights for bandwidth $w_B$ to pick the relay.

The key parameter in this method is the bucket size $k$. A large bucket will contain many relays, such that more of them are likely to have high bandwidth, but it will also make it more likely to pick a relay farther from the best path. A small bucket is less likely to have poor choices in terms of distance, but it is also less likely to include high-bandwidth relays. We choose to have $k = \lfloor\sqrt{n}\rfloor$, where $n$ is the number relays available for the specific position in the circuit. This setting puts a greater focus on distance than bandwidth. 


\subsection{Bandwidth-first}
The other way to select relays in this approach is to use bandwidth to narrow down the list of relays and then select from among these high-bandwidth relays to get low distance. In particular, we use our relay selection function with bandwidth weights $w_B$ to get a bucket of $k$ relays. We then apply our relay selection function with distance weights $w_D$ to select the relay. Here, large bucket size means potentially shorter paths but lower bandwidths, while small bucket size should yield high bandwidth choices with little optimization for distance. We again choose $k = \lfloor\sqrt{n}\rfloor$, which should ensure high bandwidth on average.

\section{Targeted Attacks in Relay Selection Adversaries}\label{targeted-attacks-whole}

The compromised circuit rates in different relay-level adversaries.
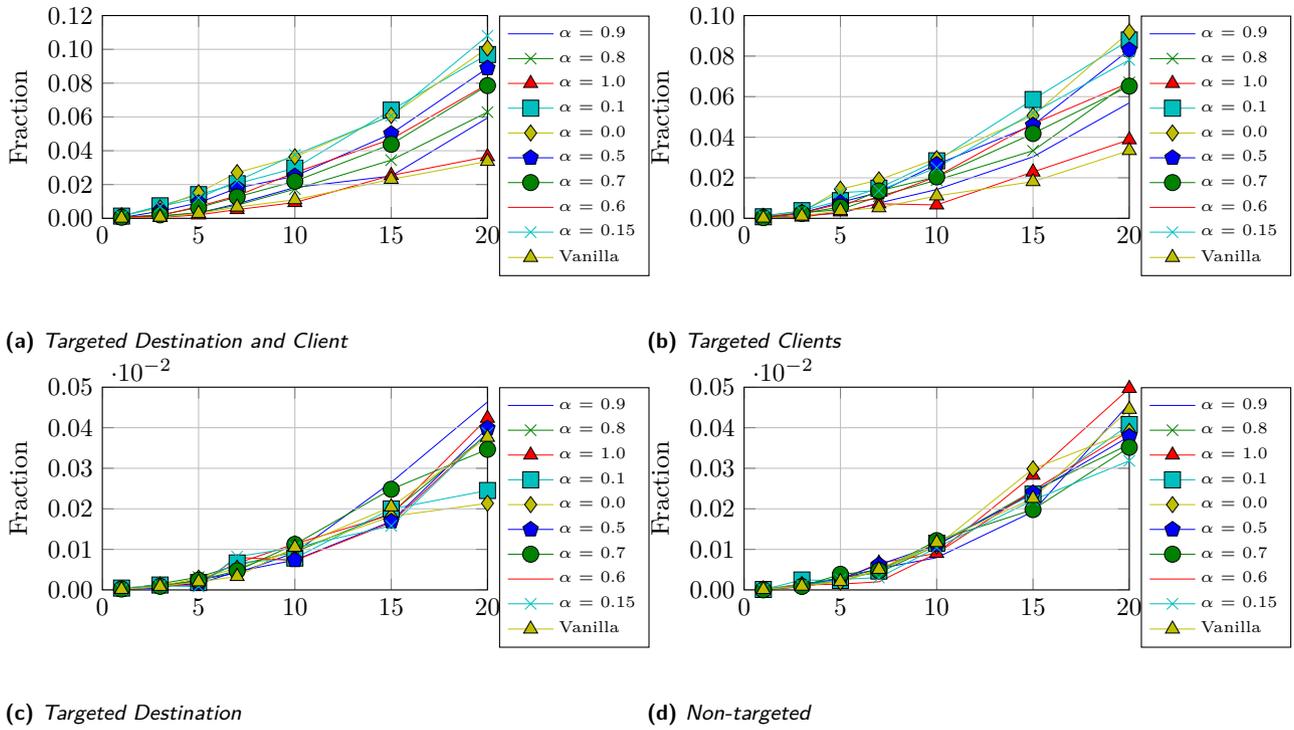
\begin{figure*}[t]
        \centering
        \begin{subfigure}[ht]{0.49\linewidth}
%
%
%
\begin{tikzpicture}

\definecolor{color1}{rgb}{0.75,0.75,0}
\definecolor{color0}{rgb}{0,0.75,0.75}

\begin{axis}[
ylabel={Fraction},
xmin=0, xmax=20,
ymin=0, ymax=0.12,
axis on top,
width=0.95\figurewidth,
height=0.85\figureheight,
xtick={0,5,10,15,20},
xticklabels={$0$,$5$,$10$,$15$,$20$},
ytick={0,0.02,0.04,0.06,0.08,0.1,0.12},
yticklabels={$0.00$,$0.02$,$0.04$,$0.06$,$0.08$,$0.10$,$0.12$},
xmajorgrids,
ymajorgrids,
legend pos=outer north east,
legend style={font=\fontsize{6}{6}\selectfont},
legend entries={{$\alpha$ = 0.9},{$\alpha$ = 0.8},{$\alpha$ = 1.0},{$\alpha$ = 0.1},{$\alpha$ = 0.0},{$\alpha$ = 0.5},{$\alpha$ = 0.7},{$\alpha$ = 0.6},{$\alpha$ = 0.15},{Vanilla}},
legend cell align={left}
]
\addplot [very thin, blue, mark=asterisk*, mark size=3, mark options={solid,draw=black}]
table {%
1 0.000234444444444444
3 0.00112
5 0.00314111111111111
7 0.00861555555555556
10 0.0182655555555556
15 0.02504
20 0.05942
};
\addplot [very thin, green!50.0!black, mark=x, mark size=3, mark options={solid}]
table {%
1 0.000667777777777778
3 0.00149222222222222
5 0.00331555555555556
7 0.00782888888888889
10 0.0172944444444444
15 0.0344222222222222
20 0.0628155555555556
};
\addplot [very thin, red, mark=triangle*, mark size=3, mark options={solid,draw=black}]
table {%
1 3.55555555555556e-05
3 0.00057
5 0.00203
7 0.00518555555555556
10 0.00934888888888889
15 0.02538
20 0.0365222222222222
};
\addplot [very thin, color0, mark=square*, mark size=3, mark options={solid,draw=black}]
table {%
1 0.00131888888888889
3 0.00741222222222222
5 0.0140355555555556
7 0.0204922222222222
10 0.0296766666666667
15 0.0640222222222222
20 0.09694
};
\addplot [very thin, color1, mark=diamond*, mark size=3, mark options={solid,draw=black}]
table {%
1 0.00153333333333333
3 0.00616444444444444
5 0.0151422222222222
7 0.02707
10 0.0364188888888889
15 0.0608277777777778
20 0.100743333333333
};
\addplot [very thin, blue, mark=pentagon*, mark size=3, mark options={solid,draw=black}]
table {%
1 0.000423333333333333
3 0.00421666666666667
5 0.00921111111111111
7 0.0178777777777778
10 0.0250033333333333
15 0.0500455555555556
20 0.0888866666666667
};
\addplot [very thin, green!50.0!black, mark=*, mark size=3, mark options={solid,draw=black}]
table {%
1 0.000626666666666667
3 0.00203555555555556
5 0.00638222222222222
7 0.0126633333333333
10 0.0219377777777778
15 0.0438288888888889
20 0.0785711111111111
};
\addplot [very thin, red, mark=asterisk*, mark size=3, mark options={solid,draw=black}]
table {%
1 0.000524444444444444
3 0.00161666666666667
5 0.00705777777777778
7 0.0133277777777778
10 0.0270811111111111
15 0.0468333333333333
20 0.0793188888888889
};
\addplot [very thin, color0, mark=x, mark size=3, mark options={solid}]
table {%
1 0.00128777777777778
3 0.00704
5 0.0120466666666667
7 0.0220844444444444
10 0.0375977777777778
15 0.06073
20 0.108028888888889
};
\addplot [very thin, color1, mark=triangle*, mark size=3, mark options={solid,draw=black}]
table {%
1 9.44444444444444e-05
3 0.000673333333333333
5 0.00291333333333333
7 0.00667333333333333
10 0.0110133333333333
15 0.0230888888888889
20 0.0337011111111111
};
\end{axis}

\end{tikzpicture}
                \caption{\textit{Targeted Destination and Client }}
                \label{fig:whole-both}
        \end{subfigure}
        \begin{subfigure}[ht]{0.49\linewidth}
%
%
%
\begin{tikzpicture}

\definecolor{color1}{rgb}{0.75,0.75,0}
\definecolor{color0}{rgb}{0,0.75,0.75}

\begin{axis}[
ylabel={Fraction},
xmin=0, xmax=20,
ymin=0, ymax=0.1,
axis on top,
width=0.95\figurewidth,
height=0.85\figureheight,
xtick={0,5,10,15,20},
xticklabels={$0$,$5$,$10$,$15$,$20$},
ytick={0,0.02,0.04,0.06,0.08,0.1,0.12},
yticklabels={$0.00$,$0.02$,$0.04$,$0.06$,$0.08$,$0.10$,},
xmajorgrids,
ymajorgrids,
legend pos=outer north east,
legend style={font=\fontsize{6}{6}\selectfont},
legend entries={{$\alpha$ = 0.9},{$\alpha$ = 0.8},{$\alpha$ = 1.0},{$\alpha$ = 0.1},{$\alpha$ = 0.0},{$\alpha$ = 0.5},{$\alpha$ = 0.7},{$\alpha$ = 0.6},{$\alpha$ = 0.15},{Vanilla}},
legend cell align={left}
]
\addplot [very thin, blue, mark=asterisk*, mark size=3, mark options={solid,draw=black}]
table {%
1 0.000163333333333333
3 0.00140222222222222
5 0.00252222222222222
7 0.00755888888888889
10 0.0142111111111111
15 0.0303277777777778
20 0.0569377777777778
};
\addplot [very thin, green!50.0!black, mark=x, mark size=3, mark options={solid}]
table {%
1 6.55555555555556e-05
3 0.00274888888888889
5 0.00430666666666667
7 0.01088
10 0.0188166666666667
15 0.0333422222222222
20 0.0670377777777778
};
\addplot [very thin, red, mark=triangle*, mark size=3, mark options={solid,draw=black}]
table {%
1 0.000187777777777778
3 0.000622222222222222
5 0.00317222222222222
7 0.00716888888888889
10 0.00661555555555556
15 0.0228644444444444
20 0.0387555555555556
};
\addplot [very thin, color0, mark=square*, mark size=3, mark options={solid,draw=black}]
table {%
1 0.000758888888888889
3 0.00377
5 0.00869666666666667
7 0.0148855555555556
10 0.0283666666666667
15 0.0585877777777778
20 0.0879511111111111
};
\addplot [very thin, color1, mark=diamond*, mark size=3, mark options={solid,draw=black}]
table {%
1 0.00102111111111111
3 0.00309444444444444
5 0.0143744444444444
7 0.0189844444444444
10 0.0294077777777778
15 0.05069
20 0.0919077777777778
};
\addplot [very thin, blue, mark=pentagon*, mark size=3, mark options={solid,draw=black}]
table {%
1 0.000535555555555556
3 0.00269888888888889
5 0.00787777777777778
7 0.01281
10 0.0266044444444444
15 0.0459211111111111
20 0.0830566666666667
};
\addplot [very thin, green!50.0!black, mark=*, mark size=3, mark options={solid,draw=black}]
table {%
1 0.000332222222222222
3 0.00242666666666667
5 0.0056
7 0.0135488888888889
10 0.0204344444444444
15 0.0418377777777778
20 0.0651866666666667
};
\addplot [very thin, red, mark=asterisk*, mark size=3, mark options={solid,draw=black}]
table {%
1 0.000375555555555556
3 0.00259
5 0.00723333333333333
7 0.00983444444444444
10 0.0206444444444444
15 0.0466222222222222
20 0.0668177777777778
};
\addplot [very thin, color0, mark=x, mark size=3, mark options={solid}]
table {%
1 0.00128888888888889
3 0.00358444444444444
5 0.0126866666666667
7 0.0136666666666667
10 0.0250022222222222
15 0.0519888888888889
20 0.0781044444444444
};
\addplot [very thin, color1, mark=triangle*, mark size=3, mark options={solid,draw=black}]
table {%
1 0.000126666666666667
3 0.00108444444444444
5 0.00378888888888889
7 0.00522333333333333
10 0.0110755555555556
15 0.01811
20 0.0334366666666667
};
\end{axis}

\end{tikzpicture}
                \caption{\textit{Targeted Clients}}
                \label{fig:whole-client-targeted}
        \end{subfigure}
        \begin{subfigure}[ht]{0.49\linewidth}
%
%
%
\begin{tikzpicture}

\definecolor{color1}{rgb}{0.75,0.75,0}
\definecolor{color0}{rgb}{0,0.75,0.75}

\begin{axis}[
ylabel={Fraction},
xmin=0, xmax=20,
ymin=0, ymax=0.05,
axis on top,
width=0.95\figurewidth,
height=0.85\figureheight,
xtick={0,5,10,15,20},
xticklabels={$0$,$5$,$10$,$15$,$20$},
ytick={0,0.01,0.02,0.03,0.04,0.05,0.06},
yticklabels={$0.00$,$0.01$,$0.02$,$0.03$,$0.04$,$0.05$,},
xmajorgrids,
ymajorgrids,
legend style={font=\fontsize{6}{6}\selectfont},
legend pos=outer north east,
legend entries={{$\alpha$ = 0.9},{$\alpha$ = 0.8},{$\alpha$ = 1.0},{$\alpha$ = 0.1},{$\alpha$ = 0.0},{$\alpha$ = 0.5},{$\alpha$ = 0.7},{$\alpha$ = 0.6},{$\alpha$ = 0.15},{Vanilla}},
legend cell align={left}
]
\addplot [very thin, blue, mark=asterisk*, mark size=3, mark options={solid,draw=black}]
table {%
1 3.55555555555556e-05
3 0.000436666666666667
5 0.00255333333333333
7 0.00415333333333333
10 0.00918777777777778
15 0.0265266666666667
20 0.0463533333333333
};
\addplot [very thin, green!50.0!black, mark=x, mark size=3, mark options={solid}]
table {%
1 4.11111111111111e-05
3 0.00152333333333333
5 0.00316555555555556
7 0.00611666666666667
10 0.00940111111111111
15 0.01903
20 0.0384
};
\addplot [very thin, red, mark=triangle*, mark size=3, mark options={solid,draw=black}]
table {%
1 0
3 0.00124555555555556
5 0.00215333333333333
7 0.00658111111111111
10 0.0113644444444444
15 0.0185466666666667
20 0.0423455555555556
};
\addplot [very thin, color0, mark=square*, mark size=3, mark options={solid,draw=black}]
table {%
1 0.000462222222222222
3 0.00121444444444444
5 0.00181777777777778
7 0.00665111111111111
10 0.00786444444444444
15 0.01997
20 0.0245166666666667
};
\addplot [very thin, color1, mark=diamond*, mark size=3, mark options={solid,draw=black}]
table {%
1 0.00037
3 0.000971111111111111
5 0.00291
7 0.00533222222222222
10 0.00991444444444444
15 0.01813
20 0.0213566666666667
};
\addplot [very thin, blue, mark=pentagon*, mark size=3, mark options={solid,draw=black}]
table {%
1 0.000142222222222222
3 0.00100666666666667
5 0.00152555555555556
7 0.00455888888888889
10 0.00735666666666667
15 0.0169077777777778
20 0.0398777777777778
};
\addplot [very thin, green!50.0!black, mark=*, mark size=3, mark options={solid,draw=black}]
table {%
1 0.00023
3 0.000897777777777778
5 0.00240333333333333
7 0.00472444444444444
10 0.0112366666666667
15 0.0248155555555556
20 0.0346744444444444
};
\addplot [very thin, red, mark=asterisk*, mark size=3, mark options={solid,draw=black}]
table {%
1 5.66666666666667e-05
3 0.00107777777777778
5 0.00106444444444444
7 0.00804333333333333
10 0.00725222222222222
15 0.0166088888888889
20 0.0387088888888889
};
\addplot [very thin, color0, mark=x, mark size=3, mark options={solid}]
table {%
1 0.000114444444444444
3 0.00100444444444444
5 0.000852222222222222
7 0.00822111111111111
10 0.0104433333333333
15 0.0158088888888889
20 0.0387388888888889
};
\addplot [very thin, color1, mark=triangle*, mark size=3, mark options={solid,draw=black}]
table {%
1 0.00013
3 0.000878888888888889
5 0.00200555555555556
7 0.00332111111111111
10 0.0105055555555556
15 0.0204855555555556
20 0.0376366666666667
};
\end{axis}

\end{tikzpicture}
                \caption{\textit{Targeted Destination}}
                \label{fig:whole-destination-targeted}
        \end{subfigure}
        \begin{subfigure}[ht]{0.49\linewidth}
%
%
%
\begin{tikzpicture}

\definecolor{color1}{rgb}{0.75,0.75,0}
\definecolor{color0}{rgb}{0,0.75,0.75}

\begin{axis}[
ylabel={Fraction},
xmin=0, xmax=20,
ymin=0, ymax=0.05,
axis on top,
width=0.95\figurewidth,
height=0.85\figureheight,
xtick={0,5,10,15,20},
xticklabels={$0$,$5$,$10$,$15$,$20$},
ytick={0,0.01,0.02,0.03,0.04,0.05,0.06},
yticklabels={$0.00$,$0.01$,$0.02$,$0.03$,$0.04$,$0.05$,},
xmajorgrids,
ymajorgrids,
legend style={font=\fontsize{6}{6}\selectfont},
legend pos=outer north east,
legend entries={{$\alpha$ = 0.9},{$\alpha$ = 0.8},{$\alpha$ = 1.0},{$\alpha$ = 0.1},{$\alpha$ = 0.0},{$\alpha$ = 0.5},{$\alpha$ = 0.7},{$\alpha$ = 0.6},{$\alpha$ = 0.15},{Vanilla}},
legend cell align={left}
]
\addplot [very thin, blue, mark=asterisk*, mark size=3, mark options={solid,draw=black}]
table {%
1 6.88888888888889e-05
3 0.00148777777777778
5 0.00276666666666667
7 0.00527777777777778
10 0.00795888888888889
15 0.0193066666666667
20 0.0459522222222222
};
\addplot [very thin, green!50.0!black, mark=x, mark size=3, mark options={solid}]
table {%
1 6e-05
3 0.00119666666666667
5 0.00306111111111111
7 0.00424111111111111
10 0.0109722222222222
15 0.0243933333333333
20 0.0359166666666667
};
\addplot [very thin, red, mark=triangle*, mark size=3, mark options={solid,draw=black}]
table {%
1 5.33333333333333e-05
3 0.000767777777777778
5 0.00233888888888889
7 0.00662111111111111
10 0.00901777777777778
15 0.0282955555555556
20 0.0496855555555556
};
\addplot [very thin, color0, mark=square*, mark size=3, mark options={solid,draw=black}]
table {%
1 0.000112222222222222
3 0.00245555555555556
5 0.00233777777777778
7 0.00466444444444444
10 0.0115277777777778
15 0.0236177777777778
20 0.0407844444444444
};
\addplot [very thin, color1, mark=diamond*, mark size=3, mark options={solid,draw=black}]
table {%
1 0.000278888888888889
3 0.00133888888888889
5 0.00198666666666667
7 0.00492888888888889
10 0.0108522222222222
15 0.02986
20 0.0390888888888889
};
\addplot [very thin, blue, mark=pentagon*, mark size=3, mark options={solid,draw=black}]
table {%
1 0.000138888888888889
3 0.000987777777777778
5 0.00281555555555556
7 0.00620888888888889
10 0.01109
15 0.02392
20 0.0378622222222222
};
\addplot [very thin, green!50.0!black, mark=*, mark size=3, mark options={solid,draw=black}]
table {%
1 2.77777777777778e-05
3 0.000862222222222222
5 0.00385888888888889
7 0.00476555555555556
10 0.0121633333333333
15 0.0198122222222222
20 0.0351622222222222
};
\addplot [very thin, red, mark=asterisk*, mark size=3, mark options={solid,draw=black}]
table {%
1 7.22222222222222e-05
3 0.00124888888888889
5 0.00143888888888889
7 0.00200555555555556
10 0.00904111111111111
15 0.0244966666666667
20 0.0391566666666667
};
\addplot [very thin, color0, mark=x, mark size=3, mark options={solid}]
table {%
1 0.00021
3 0.00104222222222222
5 0.00260333333333333
7 0.00303777777777778
10 0.0104377777777778
15 0.0221555555555556
20 0.0318277777777778
};
\addplot [very thin, color1, mark=triangle*, mark size=3, mark options={solid,draw=black}]
table {%
1 0.000104444444444444
3 0.00093
5 0.00206
7 0.00507666666666667
10 0.0116833333333333
15 0.0225688888888889
20 0.0445266666666667
};
\end{axis}

\end{tikzpicture}
                \caption{\textit{Non-targeted }}
                \label{fig:whole-non}
        \end{subfigure}        
        
        \caption{Fraction of compromised paths}\label{fig:whole-targeted-attacks}
\end{figure*}

\end{document}